%% file: main.tex
\documentclass[11pt]{article}
 
\usepackage[top=1in,bottom=1in,left=1in,right=1in]{geometry}
\usepackage[numbers]{natbib}
\input{formality.tex}
\input{ macros.tex}

\usepackage[dvipsnames]{xcolor}
\usepackage{tikz, pgfplots, pgfplotstable, subcaption}
\usepackage{hyperref}       
\hypersetup{
  colorlinks=true,
  linkcolor=black,
  urlcolor=black,
  citecolor=black
}
\usepackage{algorithm}
\usepackage{algorithmicx}
\usepackage{amsmath,amssymb,amsfonts}
\usepackage{graphicx,color}
\usepackage{url}
\usepackage{subfiles}
\usepackage{booktabs}
\usepackage{mathtools}
\usepackage{microtype}
\usepackage[english]{babel}
\usepackage{setspace}

\usepackage[figurename=Fig.,font={small,stretch=0.84}]{caption}
 
\usepackage{cleveref}
\crefname{section}{Section}{Sections}
\crefname{figure}{Figure}{Figures}
\crefname{mythm}{Theorem}{Theorems}
\crefname{mylem}{Lemma}{Lemmas}
\crefname{myrem}{Remark}{Remarks}
\crefname{appendix}{Appendix}{Appendicies}
\crefname{myprop}{Proposition}{Propositions}
\crefname{equation}{Equation}{Equations}

\usetikzlibrary{arrows.meta, calc, topaths, positioning, automata}
\pgfplotsset{compat=1.16}
\newenvironment{customlegend}[1][]{%
\begingroup
\csname pgfplots@init@cleared@structures\endcsname
\pgfplotsset{#1}%
}{%
\csname pgfplots@createlegend\endcsname
\endgroup
}%
\def\addlegendimage{\csname pgfplots@addlegendimage\endcsname}
\def\BibTeX{{\rm B\kern-.05em{\sc i\kern-.025em b}\kern-.08em
    T\kern-.1667em\lower.7ex\hbox{E}\kern-.125emX}}

\title{
  Improving Blockchain Consistency by Assigning Weights to Random Blocks
}
 
\usepackage{authblk}
\author[*]{Qing Zhang} 
\author[*]{Xueping Gong}
\author[$\ddagger$]{Huizhong Li}
\author[$\ddagger$]{Hao Wu}
\author[$\dagger$]{Jiheng Zhang}

\affil[*$\dagger$]{Department of Industrial Engineering and Decision Analytics}
\affil[$\dagger$]{Department of Mathematics}
\affil[*$\dagger$]{The Hong Kong University of Science and Technology}
\affil[$\ddagger$]{WeBank Co., Ltd.}

\date{\today}

\begin{document}
\maketitle
\renewcommand{\thefootnote}{*}
\footnotetext{Equal contributions.}

\subfile{abstract.tex}

\subfile{introduction.tex}

\subfile{model.tex}
\subfile{analysis.tex}

\subfile{ discussion.tex}

\subfile{ numerical.tex}

\subfile{literature.tex}


\bibliography{ref}

\appendix
\newpage

\subfile{appendix.tex}

\end{document}

%% file: formality.tex
\usepackage{hyperref}       
\hypersetup{
  colorlinks=true,
  linkcolor=red,
  urlcolor=magenta,
  citecolor=blue
}
\usepackage{algorithm}
\usepackage{algorithmicx}
\usepackage{amsmath,amssymb,amsfonts}
\usepackage{graphicx,color}
\usepackage{url}
\usepackage{subfiles}
\usepackage{booktabs}
\usepackage{mathtools}
\usepackage{microtype}
\usepackage[english]{babel}
\usepackage{caption}
\usepackage[toc,page,title,titletoc]{appendix}

%% file: macros.tex
\newcommand{\id}[1]{{\mathbb{I}}_{\{{#1}\}}} 

\usepackage[normalem]{ulem}

\newcommand{\prob}[1]{\mathbb{P}( #1 )}
\newcommand{\Prob}[1]{\mathbb{P}\Big( #1 \Big)}
\newcommand{\e}[1]{\mathbb{E} [ #1 ]}

\newcommand{\myto}[1]{ \xrightarrow{#1} }
\newcommand{\mynorm}[1]{ \left\|   #1  \right\| }

\newcommand{\chain}[0]{ \mathcal{C}}

\newcommand{\idc}[1]{{\bf{1}}_{\{{#1}\}}} 

\newcommand{\closure}[1]{\overline{{   F_{#1 -\Delta  }  }  }   }
\newcommand{\diff}[1]{\mathcal{W}(#1)}

\newcommand{\BB}{iron block}
\newcommand{\BBs}{iron block } 
\newcommand{\SB}{regular block}
\newcommand{\SBs}{regular block }

\newcommand{\protocol}{Ironclad}
\newcommand{\protocols}{Ironclad }

\newcommand{\qname}{randomness parameter}
\newcommand{\qnames}{randomness parameter }

\newcommand{\Dfork}[1]{\mathcal{F}_{\Delta}(#1) }
\newcommand{\diffFC}[1]{\mathcal{W}(\closure{#1})}

\newcommand{\alphaname}{minimal mining power requirement}

\newcommand{\alphanames}{minimal mining power requirement }

\def \hon {RoyalBlue}
\def \honB {RoyalBlue!15}
\def \adv {purple}
\def \advB {purple!20}

\def \colorironl {ForestGreen}
\def \colorironw {RoyalBlue}
\def \colornakal {orange}

\def \markironl {square}
\def \markironw {o}
\def \marknakal {triangle}
\def \marknine {x}
\def \markfive {*}

\newtheorem{mythm}{\bf{Theorem}}[section]
\newtheorem{mylem}{\bf{Lemma}}[section]
\newtheorem{myprop}{\bf{Proposition}}[section]
\newtheorem{mydef}{\bf{Definition}}[section]

\newtheorem{myrem}{\bf{Remark}}[section]
\newenvironment{myproof}{ {\noindent\it Proof.\ }}{\hfill $\square$\par}

%% file: abstract.tex
\begin{abstract}
Blockchains based on the celebrated Nakamoto consensus protocol have shown promise in several applications, including cryptocurrencies. 
However, these blockchains have inherent scalability limits caused by the protocol's consensus properties.
In particular, the \emph{consistency} property demonstrates a tight trade-off between block production speed and the system's security in terms of resisting adversarial attacks. 
This paper proposes a novel method, Ironclad, that improves blockchain consistency bound by assigning a different weight to randomly selected blocks.
We apply our method to the original Nakamoto protocol and rigorously prove that such a combination can improve the consistency bound significantly by analyzing the fundamental consensus properties.
Such an improvement enables a much faster block production rate than the original Nakamoto protocol under the same security guarantee with the same proportion of malicious mining power (see \cref{fig:final result}).
\end{abstract}

%% file: introduction.tex
\section{Introduction}
In 2008, Satoshi Nakamoto \cite{nakamoto2012bitcoin} proposed the celebrated Nakamoto protocol, which uses \emph{proofs of work} (PoW) to solve consensus problems in distributed systems.
Such a protocol is \emph{permissionless} since any party can participate in building such a blockchain by solving cryptographic puzzles and broadcasting valid blocks via a peer-to-peer network. 
The Nakamoto consensus protocol is proven to be highly secure as it has been widely used in applications such as cryptocurrencies in the past decade.

However, Nakamoto consensus by design has a tight trade-off between throughput and security, preventing its blockchain systems from being scaled up for a broader range of applications. 
It is well known that Bitcoin suffers from low throughput, measured by transactions per second, and long latency since the average time between two blocks is ten minutes. 
Such problems cannot be solved by simply increasing the speed of producing blocks because of the network delay and the adversary.

The scalability of blockchain systems based on the Nakamoto protocol is essentially limited by the protocol's fundamental properties, namely the agreement among honest parties, the growth rate of the longest chain, the proportion of blocks mined by honest parties \cite{backbone_protocol, asynchronous_networks, tighter_bound, better_method}.
The agreement among honest parties, referred to as \emph{consistency}, is the most crucial one 
because the \emph{consistency bound} captures the security threshold of the system \cite{Nakamoto-Win,tighter_bound}.
Therefore, improving the consistency bound allows us to increase the speed of producing blocks while tolerating the same proportion of adversarial mining power.

To analyze the consistency, the original Nakamoto paper \cite{nakamoto2012bitcoin} uses a simple random walk model to show that the probability of building a fork to exceed the longest chain drops with exponential rate.
However, the analysis does not consider other attacks where the adversary can confuse the honest parties and consequently divide the computational power owned by honest parties.
The paper \cite{backbone_protocol} provides the first rigorous study that uses the \emph{common prefix} to model and analyze the consistency in the synchronous setting.
Later, \cite{asynchronous_networks} proposes a different model to establish the consistency bound in networks with delay upper bound $\Delta$.
For such $\Delta$-synchronous networks, \cite{better_method} tightens the bound based on a Markov model;
\cite{ren2019analysis} adopts a Poisson model to obtain the consistency condition;
\cite{tighter_bound} leverages the margin analysis and random walk to establish a tight consistency bound,
and \cite{Nakamoto-Win} uses a different tree model analysis to obtain the same bound.

Many recent research efforts have committed to designing effective protocols to improve throughput based on the Nakamoto consensus. 
As summarized in the survey papers \cite{SoK_Consensus,SoK_DAG}, 
some protocols exploit blocks that are not on the main chain, achieving higher transaction processing rates \cite{ghost,GHOST_analysis,GHOST2};
some protocols replace the original chain structures with a Directed Acyclic Graph (DAG) or parallel chains consisting of concurrent chain instances \cite{GHOSTDAG,EPIC,conflux,Prism,OHIE,Chainweb};
off-chain approaches \cite{lightning-network} allow parties to execute transactions off the main chain consensus and only submit the final state to the blockchain.
In addition to the methods above, some recent works propose the decoupled consensus \cite{FruitChain,Prism,BitcoinNG,Subchains,EPIC} that separates the storage and consensus functions of a block to allow a higher rate of generating blocks that are only related to transactions.
Despite the novelty behinds these new designs, we note that many of them still rely on the original Nakamoto consensus without any fundamental improvement on the consistency.
For example, the DAG structure in \cite{EPIC} is a convergence structure that relies on a Nakamoto chain (consisting of higher difficulty blocks) to build the global ledger.

\par\smallskip
\noindent \textbf{Our approach.}
In this paper, instead of developing new protocols, we aim to improve the consistency bound of the original Nakamoto consensus. 
Note that the original Nakamoto protocol treats all valid blocks equally for both consensus and information storage.
As an upgrade, we 
randomly choose some valid blocks to be \emph{\BB s} with probability $q$. 
To differentiate, we call the unchosen valid blocks \emph{\SB s}. 
The key is to let the two types of blocks play roles of differentiated importance in the process of reaching consensus.
To this end, we introduce the concept of \emph{weight} and assign \SB s weight 1 and \BB s a larger weight $\theta>1$. 
The weight of a chain is defined to be the sum of the weights of all blocks (iron or regular) on this chain. 
Such an upgrade requires a very minimal change of the original Nakamoto protocol since what an honest party needs to do is to ``stick'' with the \emph{heaviest} chain instead of the \emph{longest} chain.
We call our method Ironclad, which is characterized by the weight and randomness parameters $(\theta,q)$.

\par
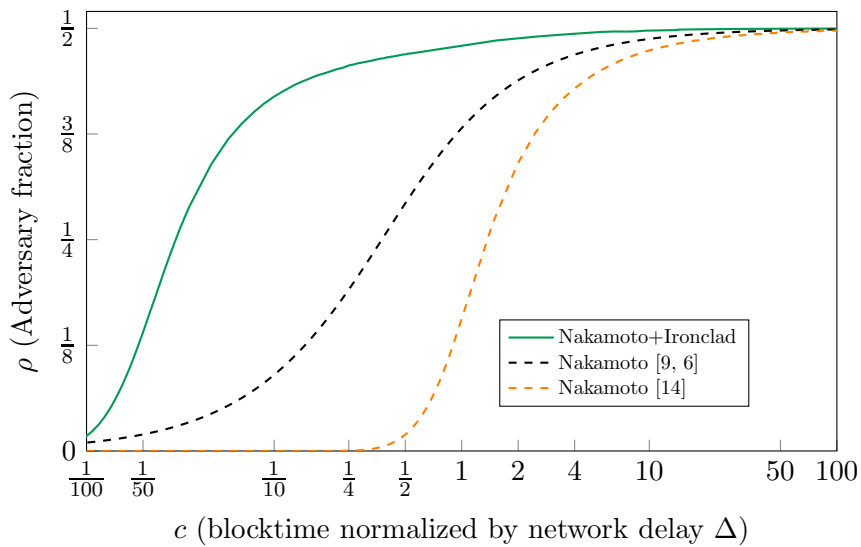
\begin{figure}[hbtp] 
\centering
\input{tex_figure/adv_hon_ratio.tex}
\caption{Minimum percentage of computing power the adversary must control in order to break consistency.
We compare the consistency bounds of Nakamoto from \cite{better_method,tighter_bound,Nakamoto-Win} with the consistency bound of Nakamoto+Ironclad established in Theorem~\ref{thm: Consistency} with $q=0.02$ and $\theta=500$. The network delay $\Delta=10^{13}$ and total mining rate $p=\frac{1}{c\Delta}$.}
\label{fig:final result}
\end{figure}

\smallskip
\noindent \textbf{Efficacy.}
Such a small change turns out to be an effective modification of the Nakamoto consensus protocol that can achieve a great improvement on the consistency bound.
To demonstrate the efficacy, \cref{fig:final result} shows a comparison of the regions where the consistency property holds in terms of the fraction of mining power $\rho$ owned by the adversary and the blocktime $c$ normalized by the network delay bound $\Delta$. 
The larger the area under the curve, the better the consistency bound, because for the same block production speed, Nakamoto consensus combined with \protocols can tolerate a larger adversarial fraction; 
for the same adversary fraction, our method allows a much higher speed of producing blocks.
The improvement is especially significant when blocks are produced at high speed (small $c$), which is in great need in practice.
%
%
Furthermore, numerical results in \cref{section: Numerical Experiments} demonstrate that in addition to the consistency, our method also improves other fundamental consensus properties such as chain quality and confirmation time.

\par\smallskip
\noindent \textbf{Intuition.}
With a large weight $\theta$, an \BBs mined by an honest party can easily beat other competing \SB s to be accepted by all honest parties, and consequently reduce the waste of honest mining power caused by forks.
The adversary does not benefit from \BB s in the same way as the honest parties do.
When the adversary get an \BB , if he broadcasts it immediately, then forks will end as in the previous case; if withholds it for breaking agreement later, such leading weights will be exceeded by the heaviest honest chain due to honest majority.
However, a too-large weight parameter $\theta$ will take too long for block confirmation, 
forming a trade-off between the confirmation time and the consistency bound illustrated in \cref{section: Discussion}.
The \qnames $q$ controls the frequency of \BB s, which could be optimized to minimize the side effects of $\theta$.
In \cref{subsec: selection of q}, we provide principles to choose parameters $(\theta,q)$ by solving an optimization problem.

\par\smallskip
\noindent \textbf{Methodology.}
We rigorously prove that our method can significantly improve the consistency bound of Nakamoto consensus.
We model the mining process by adopting the formal language theory and borrowing the concept of characteristic strings in \cite{tighter_bound}. 
We make a more precise approximation of the Poisson process to describe block production.
\cite{fork_org} uses a graph-theoretic tool to describe forks in the blockchain system, and then \cite{tighter_bound} introduces the concept called $\Delta$-fork based on the tool. 
We adopt these tools to our method with two types of blocks.
Then, we leverage the semi-Markov chain and concentration inequalities to analyze the \emph{consistency} and \emph{liveness} properties of Nakamoto+Ironclad. 
Compared with the result in \cite{better_method}, our analysis establishes a more rigorous characterization of state transitions for the more general model induced by \protocol. 
We build a two-dimensional random walk model to analyze the consensus attack and prove that Nakamoto+\protocols can outperform the Nakamoto protocol in terms of expected adversarial fork length.

\par\smallskip
\noindent \textbf{Main contributions.}
The contributions of this work can be summarized as the following. 
\begin{enumerate}
    \item We propose Ironclad, an effective and simple stochastic method to improve the blockchain consistency bound.
    \item We apply our method to the original Nakamoto protocol and rigorously prove that such combination can improve the consistency bound significantly.
    \item Furthermore, we provide principles to choose proper parameters of \protocol\ by analyzing the trade-off between the consistency bound and the confirmation time and optimizing tolerance to the adversary. 
    \item We apply our method to parallel chains and show the efficacy in the numerical experiments.
    We also discuss the potential extension to other Nakamoto-based protocols.
\end{enumerate}

\par\smallskip
\noindent \textbf{Outline.}
In \cref{section: Model}, we introduce the basic setting and related concepts describing the proposed method. 
In \cref{section: Analysis}, we first build a semi-Markov model to analyze the impact of our method on system consensus. 
Moreover, we prove that our method enhances consensus for any non-trivial \qname\ $q\in (0,1)$ and $\theta>1$.
Additionally, we discuss the selection of relevant parameters in our approach. 
We also show that the upgraded protocol satisfies chain growth and chain quality. 
In \cref{section: Discussion}, a random walk model is built to analyze attacks in a detailed way. 
Finally, we apply our idea to Nakamoto protocol and parallel chains in \cref{section: Numerical Experiments}, 
and obtain convincing simulation data to verify our conclusion.
We review literatures and discuss potential combinations with our method in \cref{sec:literature}.


\par\smallskip
\noindent \textbf{Notations.}
To facilitate the description of our model, we first present a set of notations. 
For a natural number $n$, $[n]$ denotes the set $\{1,2,\cdots,n\}$. 
An indicator function of an event $E$ is denoted by $\idc{E}$. 
Let $\Sigma$ be an alphabet containing several symbols, e.g., $\Sigma=\{0,a,A,h,H\}$.
Denote $\Sigma^n$ as the set of all strings consisting of symbols in $\Sigma$ with length $n$ 
and $\Sigma^* = \Sigma^0\cup \Sigma^1\cup\cdots \cup \Sigma^n\cup\cdots$ as the set of all such strings of any length.
For any string $w$, denote $|w|$ its length.
A language is a subset of $\Sigma^*$ and the concatenation of two languages $L_1$ and $L_2$ is denoted by 
$L_1||L_2 = \{ w w' | w\in L_1, w'\in L_2\}$ where $ww'$ is the concatenation of two strings.
In some cases, we are also interested in the slice of a string $w=w_1w_2\cdots$, 
where $w_i$ is the i-th symbol in $w$.
Denote $w_{s:t}=w_sw_{s+1}\cdots w_t$ as the slice of $w$ from position $s$ to $t$.

%% file: tex_figure/adv_hon_ratio.tex
\begin{tikzpicture}
\begin{semilogxaxis}[
    width=0.7\textwidth,
    height=0.45\textwidth,
    ylabel = $\rho$ (Adversary fraction),
    xlabel = $c$ (blocktime normalized by network delay $\Delta$),
    log basis x = 2 ,
    ymin=0, ymax=0.52,
    xmin=0.01, xmax=100,
    xtick pos = left,
    ytick pos = left,
    xtick={0.01,0.02,0.1,0.25,0.5,1,2,4,10, 50, 100},
    xticklabels ={$\frac{1}{100}$,$\frac{1}{50}$,$\frac{1}{10}$,$\frac{1}{4}$,$\frac{1}{2}$,1,2,4, 10, 50, 100},
    ytick={0,0.125,0.25,0.375,0.5},
    yticklabels ={0,$\frac{1}{8}$,$\frac{1}{4}$,$\frac{3}{ 8}$,$\frac{1}{2}$},
    legend cell align={left},
    legend style={at={(0.55,0.2)}, anchor=west, nodes={scale=0.7, transform shape}},
]
\addplot[mark=none, \colorironl, thick] table [x index=0,y index=1,col sep=space] {tex_figure/adv_C_H_i.dat};
\addplot[mark=none, dashed, thick] table [x index=0,y index=1,col sep=space] {tex_figure/tight_bound.dat};
\addplot[mark=none, \colornakal, dashed,thick] table [x index=0,y index=1,col sep=space] {tex_figure/adv_C_h.dat};

\legend{
    Nakamoto+Ironclad,
    Nakamoto \cite{tighter_bound,Nakamoto-Win},
    Nakamoto \cite{better_method},
}
\end{semilogxaxis}
\end{tikzpicture}

%% file: model.tex
\section{Problem Formulations}
\label{section: Model}

We consider a $\Delta$-synchronous, slot-based and permissionless network model.
A blockchain system is maintained by a number of parties, including the honest and the adversary.
Honest parties carry out the specified protocol, while the adversary may diverge arbitrarily from the specifications.
More precisely, honest parties keep a copy of their current view of the blockchain, try to mine blocks at the end of their chain,
and broadcast the result to all other parties immediately.
Adversarial parties can observe messages as soon as it is sent. 
They can delay and reorder all messages received by honest parties up to a delay of $\Delta$ time slots. 
All parties actively engage in searching for ``proofs-of-work” (PoWs) by accessing a random oracle.

We combine our method with the original Nakamoto protocol \cite{backbone_protocol,asynchronous_networks,better_method,tighter_bound} 
by randomly selecting blocks to be \BB s and assigning weights to them.
To this end, we introduce the concept of \emph{weight} and assign \SB s weight 1 and \BB s a larger weight $\theta>1$. 
The weight of a chain is defined to be the sum of the weights of all blocks (iron or regular) on this chain (see \cref{eq:weight}) . 
Such an upgrade requires a very minimal change of the original Nakamoto protocol since what each honest party needs to do is to ``stick'' with the \emph{heaviest} chain instead of the \emph{longest} chain.

\subsection{Characteristic String}
\label{subsection:model}

For the purposes of analysis, we discretize the continuous timeline into time slots indexed by natural numbers.
In order to analyze the dynamics of block production given the presence of the adversary and network delay, we adopt the models from \cite{tighter_bound,fork_org} and extend the analysis to our case with two types of blocks.
We use a \emph{characteristic symbol} to indicate whether a proof-of-work is discovered in a particular time slot,  and whether the successful party is honest or adversarial.
A symbol from the alphabets $\{0,h,H\}$ and $\{0,a,A\}$ indicates the mining outcome in each time slot by the honest parties and the adversary, respectively.
Specifically, $0$ means no block produced, $\{h,a\}$ means a \SB, and $\{H,A\}$ means an \BB.
A sequence of symbols in $\{0,h,H\}$ and $\{0,a,A\}$ consists of a \emph{characteristic string} for honest and adversarial parties, respectively.

The probability that more than one block can be produced in a time slot diminishes as the discretization of the continuous timeline becomes finer.
Thus, we can exclude such an event and assume that at most one block can be produced in each time slot \cite{better_method,tighter_bound}. 
Let $p$ be the probability of obtaining a block in a time slot and $\rho$ be the adversarial fraction of total mining power. 
Then we define  $p_h\coloneqq (1-\rho)p$ and $p_a \coloneqq \rho p$ to be the probabilities of obtaining a block in a time slot by the honest parties and the adversary, respectively,
and $q$ to be the probability for a newly mined block to be an \BB.
It is clear that for any given $p_h,p_a,q$, each symbol in honest characteristic strings is independently drawn according to
\begin{equation*}
w_t = \left\{
\begin{array}{ll}
h, &\text{with probability } (1-q)p_h,\\
H, &\text{with probability } qp_h,\\
0, &\text{with probability } 1-p_h.
\end{array}
\right.
\end{equation*}
Similarly, each symbol in the adversary characteristic string $w'_1w'_2\cdots\in\{0,a,A\}^*$ is independently drawn according to
\begin{equation*}
w'_t = \left\{
\begin{array}{lll}
a, &\text{with probability } (1-q)p_a,\\
A, &\text{with probability } qp_a,\\
0, &\text{with probability } 1-p_a.
\end{array}
\right.
\end{equation*}


Note that blocks discovered by the honest parties and the adversary are mutually exclusive in \cite{tighter_bound}, while the behaviors of the two parties are independent in our model, closer modeling to reality.
Therefore, it is possible to have at most two blocks (one from the honest and one from the adversary) in a time slot. 

\begin{myrem}
	The selection of \BB s requires the consensus guarantee, which means that anyone is able to verify the validity of \BB s.
	In general, verifiable random variables (e.g., block hash) with a known distribution can satisfy the requirement.
	We provide several methods that are widely used and easy to implement.
	A classical and natural way is to set difficulty targets and compare them with block hashes, 
	which is introduced by Bitcoin\cite{nakamoto2012bitcoin} and widely used in literature \cite{Proofs_of_proofs_of_work,OHIE, Prism}.
	A direct extension is rehashing via a new independent random oracle. 
	As introduced by \cite{backbone_protocol}, 2-for-1 trick can also be used to differentiate blocks:
	the prefix of a block hash determines whether the mining is successful, and the suffix is used to determine block types \cite{FruitChain}.	
\end{myrem}

\subsection{Blockchain Structure and Our Goals}
\label{subsec: forks}

Since all valid blocks form a directed rooted tree $F=(V,E)$, where $V$ is the set of vertices (blocks) and $E$, it is also important to characterize the graphic structure of \SB s and \BB s.
For any vertex $B\in V$, define $depth(B)$ to be the length of the shortest path from the root to $B$.
We define a label function $lb: V \to \{0,1,2,\ldots\}$ to map a block $B\in V$ to the time slot when it was produced. 
A block $B$ is honest if it is mined by an honest party, and adversarial otherwise.
Using the notation of characteristic string and label function., if block $B$ is honest then $w_{lb(B)}\in\{h,H\}$.
However, there may exist an adversarial block $B'$ produced in the same time slot, i.e., $lb(B')=lb(B)$ but $w_{lb(B)}\in\{a,A\}$.
The tool to analyze the graphic structure originates from \cite{fork_org}. 
\cite{asynchronous_networks} consider the delay in a tree-based model and \cite{tighter_bound} systematically describe the $\Delta$-fork via characteristic strings. 
We reflect network delays with a single parameter $\Delta$:
while any message sent by honest parties will be delivered, 
the adversary may delay its arrival by up to $\Delta$ time slots.
We modify the definition of $\Delta$-fork in our model for the purpose of analyzing \protocol.

\begin{mydef}[$\Delta$-fork]
\label{def:fork}
A PoW $\Delta$-fork for the honest string $w\in\{0, h, H\}^*$ and the adversarial string $w'\in\{0, a, A\}^*$ with equal length is a directed rooted tree $F=(V,E)$ satisfying the following axioms: 
\begin{enumerate}
    \item[A1.] The root $r$, which has label $lb(r)=0$, is honest.
    \item[A2.] The sequence of labels $lb(\cdot)$ along any directed path is increasing.
    \item[A3.] If $w_t\in\{h,H\}$ and $w'_t=0$ then there is exactly one block with label $t$. 
    If $w_t\in\{h,H\}$ and $w'_t\in\{a,A\}$ then there exists at least one block with label $t$.
    \item[A4.] For any two honest blocks $B_1$ and $B_2$ with $lb(B_1) + \Delta \leq lb(B_2)$, we have $depth(B_1) \leq depth(B_2)$.
\end{enumerate}
\end{mydef}

\begin{myrem}
A.1 rules out the trivial case where the honest miners and the adversary mine blocks in two completely different blockchains.
A.2 excludes the event of collisions, which has been proven to have negligible probability \cite{backbone_protocol}.
It can be inferred from A.3 that there are at most two blocks with the same label.
And if two blocks in $F$ have the same label $t$, then one must be $w_t\in\{h,H\}$ and the other must be $w'_t\in\{a,A\}$. 
This echos our model that honest parties and the adversary produce blocks independently instead of exclusively. 
A.4 means that all honest parties will take into account any honest block produced $\Delta$ (upper bound of network delay) time ago due to sufficient time for network propagation.
\end{myrem}

Denote the collection of all $\Delta$-forks for $(w,w')$ as $\mathcal{F}_{\Delta}(w,w')$. 
A path $\mathcal{C}$ in a fork $F\in\mathcal{F}_{\Delta}(w,w')$ originating at the root is called a chain. 
If the last block in $\chain$ is honest, then $\chain$ is an honest chain, indicating that $\chain$ will be held by at least one honest party.
Define 
\begin{equation}
	\label{eq:weight}
  \diff{\chain} 
  = \sum_{B\in\chain}\idc{B\text{ is a \SB }}+\theta\idc{B\text{ is an \BB }}
\end{equation}
to be the sum of the weights of all blocks in $\chain$.
Note that in the original Nakamoto protocol, the $\diff{\chain}$ degenerates to be the number of blocks in the chain $\chain$ since all blocks have the same weight. 
\protocols requires honest parties to mine a new block extending the heaviest chain in a $\Delta$-fork $F$, whose weight is defined to be the weight of its heaviest chain, i.e.,
\begin{equation}
	\label{eq:weight-fork}
  \diff{F} = \max_{\chain\subset F} \diff{\chain}.
\end{equation}


\begin{mydef}[Honest Subfork]
A subfork $F'\subset F\in\mathcal{F}_{\Delta}(w,w')$ is honest if all the leaves of $F'$ are honest.
The honest subfork with the most blocks in $F$ is the maximal honest subfork, denoted as $\overline{F}$.
\end{mydef}

In many cases, we are interested in the chains and forks up to time slot $t$. 
We add subscripts $t$ to chains and forks to emphasize that they are created by the time slot $t$.
Intuitively, $F_t$ and $\chain_t$ are ``snapshots'' of forks and chains at time $t$.
For example, $F_t$ is an element in $\Dfork{w_{1:t},w'_{1:t}}$, and $\chain_t\subset F_t$.
In order to take the delay $\Delta$ into consideration, we need to remove the most recent $\Delta$ time slots from time $t$.
For example, the $\Delta$-fork $F_t$ without considering the latest $\Delta$ time slots can be represented by $F_{t-\Delta}$.
We define $\chain_t\subset F_t$ to be a \emph{dominant chain} in the fork $F_t$ if 
\begin{equation*}
	\diff{\chain_t} \geq \diff{ \overline{{F}_{t-\Delta}} }.	
\end{equation*}
A dominant chain is a candidate for honest parties to follow due to the delay because no chain held by honest parties at time slot $t-\Delta$ is heavier than it.

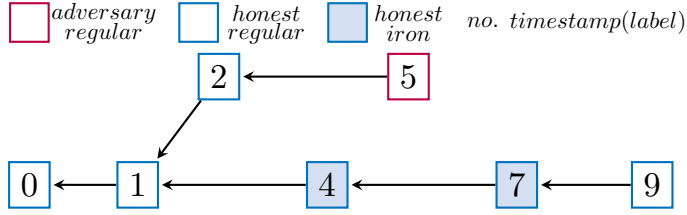
\begin{figure}[ht]
\centering
\begin{tikzpicture}[->,shorten >=1pt,auto,node distance=1cm,-stealth,
			thick,base node/.style={square,draw ,minimum size=48pt}, real node/.style={double,square,draw,minimum size=50pt},scale = 1.8]
	\node [draw=\adv,  scale = 2] (0)at(0,1.2){$ $};
	\node [scale = 0.8] (0)at(0.55,1.27){$adversary$};
	\node [scale = 0.8] (0)at(0.55,1.12){$regular$};
	\node [draw=\hon, scale = 2] (0)at(1.25,1.2){$ $};
	\node [scale = 0.8] (0)at(1.75,1.27){$honest$};
	\node [scale = 0.8] (0)at(1.75,1.12){$regular$};
	\node [draw=\hon, fill=\honB,  scale = 2] (0)at(2.35,1.2){$ $};
	\node [scale = 0.8] (0)at(2.8,1.27){$honest$};
	\node [scale = 0.8] (0)at(2.8,1.12){$iron$};
	\node [scale = 0.8] (0)at(3.35,1.2){$no.$};
	\node [scale = 0.8] (0)at(4.2,1.2){$timestamp(label)$};

	\node [draw=\hon, scale = 1.2 ] (0)at(0,0){$0$};
	\node [draw=\hon, scale = 1.2 ] (1) at(0.8,0){$1$};
	\node [draw=\hon, fill=\honB , scale = 1.2 ] (4)  at(2.2,0) {$4$};
	\node [draw=\hon , scale = 1.2  ] (2)  at(1.4,0.8) {$2$};
	\node [draw=\hon, fill=\honB , scale = 1.2 ] at(3.6,0) (7){$7$};
	\node [draw=\adv, scale = 1.2 ] (5) at(2.8,0.8){$5$};
	\node [draw=\hon, scale = 1.2 ] (9) at(4.6,0){$9$};
	\path[]
		(1) edge node {} (0)
		(2) edge node {} (1)
		(4) edge node {} (1)
		(5) edge node {} (2)
		(7) edge node {} (4)
		(9) edge node {} (7);
\end{tikzpicture}
\caption{The delay $\Delta=2$, $t=10$, honest string $w=hh0H00H0h$ and adversarial string $w'=0000a0000$.
}
\label{fig:example for Delta-fork}
\end{figure}

We now provide an example depicted in \cref{fig:example for Delta-fork} to illustrate the concepts defined above.
The fork $F_8$ exclude the block labeled $9$ and the fork $F_{8 -{\Delta}  }$ exclude the block labeled $7$. 
The block labeled $5$ is not in the maximal honest subfork $\overline{F_{8 -{\Delta}  }}$, since it is adversarial.
For the dominant chains, we have $\chain_7 =\chain_8 \subset \chain_9$.
Dominant chains in $F_9$ are $ 0\leftarrow1\leftarrow4\leftarrow7\leftarrow9 $ and $0\leftarrow1\leftarrow4\leftarrow7$, 
where the numbers are labels of corresponding blocks.
%

Based on previous formulations, we study the two properties: \emph{consistency} and \emph{liveness},  originally formulated in \cite{backbone_protocol}, 
when combining \protocols with original Nakamoto consensus.
We first introduce the definitions of chain prefix~\cite{tighter_bound}: 
For a chain $\chain_{t}$ and a natural number $k$, 
$\chain_{t-k}$ is the chain obtained by removing from $\chain_t$ all blocks minded from slots $\{t-k+1,\dots,t\}$.
When $t \leq k$, define $\chain_{t_1 - k }$ as the genesis.
Define $\chain^\prime_{t^\prime} \preceq \chain_t$ iff $\chain^\prime_{t^\prime} = \chain_{t-k}$ for some $k$,
and we say $\chain^\prime_{t^\prime}$ is a prefix of $\chain_t$.

\begin{mydef}[Consistency]
	\label{def:consistency}
	For a $\Delta$-fork $F_t$, there exists a time interval parameter $k$ such that for any slot $t_1<t_2$, 
	and dominant chains $\chain_{t_1}$ and $\chain_{t_2}$ both held by honest parties, $\chain_{t_1 - k } \preceq \chain_{t_2}$ with probability $1-\exp(-\Omega(k))$.

\end{mydef}

\begin{mydef}[Liveness]
	\label{def:liveness}
	For any slots $t_1,t_2$ with $t_1+u\leq t_2$, and any dominant chain $\chain_{t_2} \subset F_{t_2}$, 
	there exists an honest chain $\chain_{t} \preceq \chain_{t_2}$ for $t_1\leq t\leq t_1+u$  with probability $1-\exp(-\Omega(u))$.	
\end{mydef}

Intuitively, consistency means that honest parties will all agree upon a chain among all the blocks that have been fully synchronized, 
i.e., produced $\Delta$ time slots ago, with large probabilities. 
Liveness guarantees that the blockchain held by an honest party incorporates at least one fresh honest block over any period of $u$ slots.  

%% file: analysis.tex
\section{Analysis on Blockchain Properties}
\label{section: Analysis}

Consistency is the most crucial property as it guarantees that all honest parties agree on a chain except for the blocks produced in the latest $k$ time slots.
In \cref{subsec: consistency}, we analyze the consistency of Nakamoto protocol equipped with \protocols using a semi-Markov framework. 
We compare our consistency bound with the result for the original Nakamoto protocol which is established in \cite{better_method,tighter_bound,Nakamoto-Win}. 
Consistency also demonstrates a tight trade-off between security and throughput. 
In \cref{subsec: selection of q}, we discuss the selection of parameters in our method to optimizing the tolerance ratio for the adversary mining power.
Moreover, we also study three interesting properties of our approach, 
and validate them by simulation data.
Finally, we quantify the chain growth rate and chain quality, which describes the blockchain system in a different view.
We combine the three chain properties to prove the liveness property in \cref{subsec:growth and quality}.

\input{analysis/pattern.tex}

\input{analysis/markov.tex}

\input{analysis/comparison.tex}

\input{analysis/growth_and_quality.tex}

%% file: analysis/pattern.tex
\subsection{Consistency}
\label{subsec: consistency}
We analyze the state transitions caused by four characteristic patterns in our proposed semi-Markov chain model. 
The analysis relies on ``patterns'' that can guarantee the convergence of all honest chains if there is no attack from the adversary. 
For example, the pattern $\{0\}^\Delta\{h\}\{0\}^\Delta$ means that no two honest blocks produced within the synchronization delay bound \cite{backbone_protocol,asynchronous_networks,better_method}.
Such a pattern is conservative for establishing consistency but reasonable because they consider the worst cases due to the delay. 
When such a pattern occurs, the chains kept by honest parties have an opportunity to converge. 
To ruin this opportunity, the adversary must have a chain that is at least as long as the longest chain of honest parties. 
This requires the adversary to have enough computing power to produce more blocks than the occurrence number of such patterns, which has a very small probability due to the honest majority assumption.
To obtain a proper bound for the adversarial mining power tolerance, we introduce a new random variable $\alpha$ (see \cref{well-defined alpha}) to measure the minimal adversarial mining power requirement to 
prevent honest parties from reaching an agreement.

\begin{mydef}[Characteristic Patterns]
\label{def:pattern decomposition}
For a characteristic string $w\in \{h,H,0\}^*$, consider the decomposition $w=\sigma_{(1)}\sigma_{(2)}\cdots$ where $\sigma_{(i)}$ belongs to one of four patterns defined below:
\begin{itemize}
    \item $\sigma_1:\{h\}||\{0\}^k, k< \Delta$, corresponding to the case where two honest \SB s are mined within an interval smaller than the delay, which may cause forks.
    \item $\sigma_2:\{h\}||\{0\}^k, k\geq \Delta$, corresponding to the case where an honest \SBs is mined without a competing block within $\Delta$ time slots.
    \item $\sigma_3:\{H\}||\{0,h\}^k, k\leq \Delta$, similar to $\sigma_1$, the fork is caused by \BB s.
    \item $\sigma_4:\{H\}||\{0,h\}^\Delta || \{0\}^*$, similar to $\sigma_2$, an \BBs has no other competing \BB s within $\Delta$ time slots.
\end{itemize}
\end{mydef}

\begin{myrem}
    To guarantee that the \BBs in $\sigma_4$ can beat competing \SB s (if any), 
    we require that $\theta$ should be larger than the number of \SB s mined in $2\Delta$ time slots with high probability,
    which induces a lower bound of $\theta$, see \cref{lem: setting theta} and \cref{eq:theta lower bound}. 
\end{myrem}


Patterns $\sigma_1$ and $\sigma_3$ represent the cases where another block is produced too soon (within synchronization delay) after an honest block (regular or iron). 
Conversely, patterns $\sigma_2$ and $\sigma_4$ represent the cases where no honest block and no \BB\ is mined within $\Delta$ time slots after an honest \SB\ and an honest \BB, respectively. 
Let a characteristic string $w\in \{h,H,0\}^*$ starting with $H \text{ or } h$. 
Following the greedy pattern matching principle, the decomposition of $w$ into the above four patterns in Definition~\ref{def:pattern decomposition} uniquely exists.
We omit the proof since it is straightforward.

We can explicitly calculate the occurrence probability and expected length since each symbol is drawn independently from an identical multinomial distribution.
For notation brevity, define the probability for two types of blocks:
$$
q_a = (1-q)p_a, q_A = qp_a, 
q_h = (1-q)p_h, q_H = qp_h.
$$
The probability of the occurrence and the expected length of $\sigma_i$ can be easily calculated as 
$$
\begin{array}{lll}
 \prob{\sigma_1} = q_h[1-(1-p_h)^\Delta],  
& \e{|\sigma_1|} = \frac{1}{p_h}-\frac{\Delta(1-p_h)^{\Delta}}{1-(1-p_h)^\Delta},\\
 \prob{\sigma_2} = q_h(1-p_h)^\Delta, 
& \e{|\sigma_2|} = \frac{1}{p_h}+\Delta,\\
 \prob{\sigma_3} = q_H[1-(1-q_H)^\Delta], 
& \e{|\sigma_3|}= \frac{1}{q_H}-\frac{\Delta(1-q_H)^{\Delta}}{1-(1-q_H)^\Delta},\\
 \prob{\sigma_4} = q_H(1-q_H)^\Delta, 
& \e{|\sigma_4|} = \frac{1}{p_h}+\Delta.\\
\end{array}
$$

%% file: analysis/markov.tex
\par
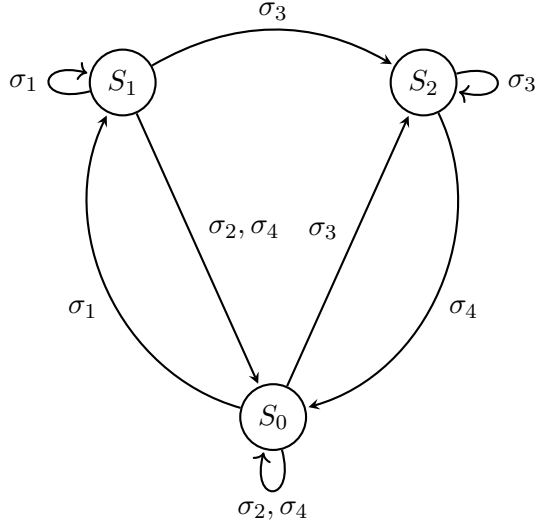
\begin{figure}[h]
\center
	\begin{tikzpicture}[->,shorten >=1pt,auto,node distance=4cm,-stealth,
			thick,base node/.style={circle,draw,minimum size=48pt}, real node/.style={double,circle,draw,minimum size=50pt},scale = 1.5]
			
	\node[shape=circle,draw=black](1){$S_1$};
			\node[shape=circle,draw=black](2)[right of=1 ]{$S_2$};
			\node[shape=circle,draw=black](0)[below = of $(1)!.5!(2)$]{$S_0$};

			\path[]
			(0) edge [loop below]node {$\sigma_2,\sigma_4$} (0)
			(0) edge [bend left=50]node {$\sigma_1$} (1)
			(0) edge node[bend right=100] {$\sigma_3$} (2)
			
			(1) edge node {$\sigma_2,\sigma_4$} (0)
			(1) edge [loop left]node {$\sigma_1$} (1)
			(1) edge [bend left]node {$\sigma_3$} (2)
			
			(2) edge [bend left=50]node {$\sigma_4$} (0)
			(2) edge [loop right]node {$\sigma_3$} (2);
            
	\end{tikzpicture}
	\caption{The state transition diagram. 
	For example, if the system is at $S_0$ and a pattern $\sigma_3$ occurs, then the Markov chain will transit into $S_1$. }
  \label{fig:state_transition_graph}
\end{figure}

\smallskip
\noindent \textbf{Analysis of the Semi-Markov Chain.}
We define three Markov states based on the four patterns defined above to capture the evolution of the blocks. 
State $S_0$ is visited whenever patterns $\sigma_2$ or $\sigma_4$ occurs, which means that honest parties have a chance to agree on the heaviest chain.
States $S_1$ and $S_2$ correspond to $\sigma_1$ and $\sigma_3$, respectively. 
In these two states, honest parties are very likely to suffer from forks due to the network delay, even in the absence of the adversary.
The transition graph is shown in \cref{fig:state_transition_graph}. 
Each edge corresponds to the pattern that causes the transition.
The transition probability from state $S_i$ to $S_j$ is denoted as $P_{ij}$.

\begin{myprop} [Markovian Property]
	The blockchain system shown in \cref{fig:state_transition_graph} is a semi-Markov chain,	and its embedded Markov chain is ergodic.
	The transition probabilities are
	\begin{equation}
		\label{transition_p}
	\begin{array}{lll}
	P_{00} = \frac{ {\prob{\sigma_2}+\prob{\sigma_4}}}{p_h},    
	& P_{01} = \frac{\prob{\sigma_1}}{p_h},   
	& P_{02} = \frac{\prob{\sigma_3}}{p_h},  \\
	P_{10} = \frac{\prob{\sigma_2}+\prob{\sigma_4}}{p_h},  
	& P_{11} = \frac{\prob{\sigma_1}}{p_h},   
	& P_{12} = \frac{\prob{\sigma_3}}{p_h}, \\
	P_{20} = \frac{\prob{\sigma_4}}{q_H},   
	& P_{21} = 0,   
	& P_{22} = \frac{\prob{\sigma_3}}{q_H}.
	\end{array}
	\end{equation}
\end{myprop}

\begin{myproof}
The blockchain system makes a state transition whenever a pattern occurs. 
Therefore, we can regard each pattern as one step.
Given the present state, the next state is independent of history and only depends on the present state.
Hence, it satisfies the Markovian property.
Obviously, the chain is irreducible and the three states are positive recurrent. 
Since the transition time between any two states is random, the blockchain system is a semi-Markov chain and 
the embedded Markov chain is ergodic.

The transition probabilities between two states are proportional to the pattern occurrence probabilities corresponding with the edge between two states.
From \cref{fig:state_transition_graph}, the transition probability in \cref{transition_p} can be computed by conditional probabilities.
For example, the transition $S_0 \to S_1$ happens when an honest block is mined and this block constitutes a pattern $\sigma_1$ with the following `0' symbols less than $\Delta$. 
Therefore, $P_{01}$ is the conditional probability of pattern $\sigma_1$ occurrence when an honest block was mined and the system was at $S_0$.
Other transition probabilities can be interpreted similarly.
\end{myproof}

Compared with the Markov model in \cite{better_method}, our model extends it in two aspects:
\begin{itemize}
\item We explicitly provide more precise definitions of states and transition events. 
The transition time between states depends on the length of patterns that are not ``memoryless" random variables, 
namely, how long the system has stayed in its current state. 

\item To characterize the effect of \BB s, we add a new state $S_2$ that is not symmetric with respect to $S_1$ due to the large weight of \BB s.
\end{itemize}

\begin{figure}[h]
	\centering
	\begin{tikzpicture}[->,shorten >=1pt,auto,node distance=1cm,-stealth,
				thick,base node/.style={square,draw ,minimum size=48pt}, real node/.style={double,square,draw,minimum size=50pt},scale = 1.7]
		

		\node [draw=\adv,  scale = 2] at(0,1.2){$ $};
		\node [scale = 0.8] at(0.6,1.27){$adversary$};
		\node [scale = 0.8] at(0.6,1.12){$regular$};
		\node [draw=\adv, fill=\advB,  scale = 2] at(1.4,1.2){$ $};
		\node [scale = 0.8] at(2.,1.27){$adversary$};
		\node [scale = 0.8] at(2.,1.12){$iron$};
		\node [draw=\hon,  scale = 2] at(2.8,1.2){$ $};
		\node [scale = 0.8] at(3.35,1.27){$honest$};
		\node [scale = 0.8] at(3.35,1.12){$regular$};
		\node [draw=\hon, fill=\honB, scale = 2] at(4,1.2){$ $};
		\node [scale = 0.8] at(4.5,1.27){$honest$};
		\node [scale = 0.8] at(4.5,1.12){$iron$};

		\node [draw=\hon, dashed, scale = 2.2] (0)at(0,-0.8){$ $};
		\node [draw=\hon,  scale = 1.2] (1) at(1.2,-0.8){$1$};
		\node [draw=\hon, fill=\honB, scale = 1.2 ] (4)  at(3.2,-0.8) {$4$};
		\node [draw=\hon ,scale = 1.2] (2)  at(2,0) {$2$};
		\node [draw=\hon, scale = 1.2,fill=\honB ] at(4.8,-0.8) (7){$7$};
		\node [draw=\hon, scale = 1.2] (5) at(3.8,0){$5$};
		\node [draw=\adv, fill=\advB, scale = 1.2] (6) at(3.2,0.8){$4$};
		
		\draw [dotted,-] (1.05,-1.35)--(1.05,1);  \node (S0) at(1.05,-1.4){$S_0$};
		\draw [dotted,-] (1.85,-1.35)--(1.85,1); \node (S1) at(1.85,-1.4){$S_1$};
		\draw [dotted,-] (3.05,-1.35)--(3.05,1); \node (S11) at(3.05,-1.4){$S_1$};
		\draw [dotted,-] (4.65,-1.35)--(4.65,1); \node (S01) at(4.65,-1.4){$S_0$};
			\path[]
				(1) edge node {} (0)
				(2) edge node {} (0)
				(4) edge node {} (1)
				(5) edge node {} (2)
				(6) edge node {} (2)
				(7) edge node {} (4)
				(S0) edge node {$\sigma_1$} (S1)
				(S1) edge node {$\sigma_1$} (S11)
				(S11) edge node {$\sigma_4$} (S01)
				;
	\end{tikzpicture}
	\caption{The convergence due to an honest \BB. 
	The $\Delta$-fork is based on $w=\cdots hh0Hh0H$ with $\Delta=2, \theta=4$.
	}
	\label{fig: high-diff block}
\end{figure}
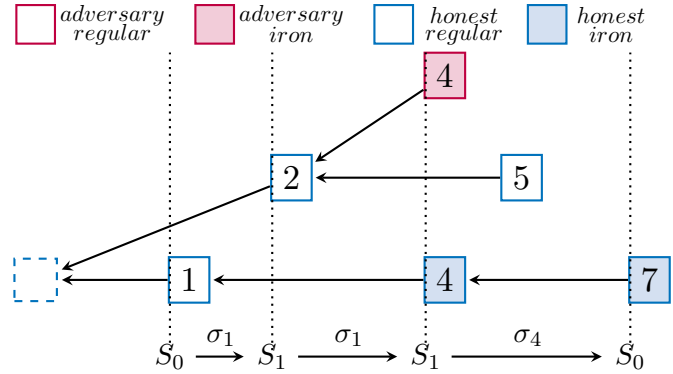
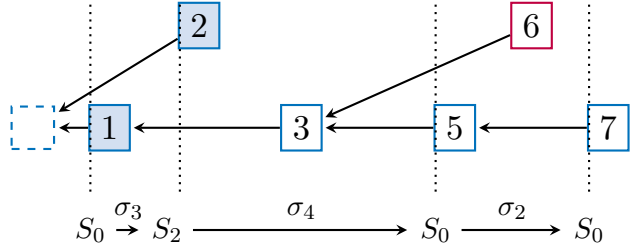
\begin{figure}[h]
	\centering
	\begin{tikzpicture}[->,shorten >=1pt,auto,node distance=1cm,-stealth,
				thick,base node/.style={square,draw ,minimum size=48pt}, real node/.style={double,square,draw,minimum size=50pt},scale = 1.7]

		\node [draw=\hon, dashed, scale = 2.2] (0)at(0.1,0){$ $};
		\node [draw=\hon, fill=\honB, scale=1.2] (1) at(0.7,0){$1$};
		\node [draw=\hon, fill=\honB, scale=1.2] (2)  at(1.4,0.8) {$2$};
		\node [draw=\hon, scale=1.2 ] (3) at(2.2,0){$3$};
		\node [draw=\hon,  scale=1.2 ] (5) at(3.4,0){$5$};
		\node [draw=\hon, scale=1.2 ] (7) at(4.6,0){$7$};
		\node [draw=\adv, scale = 1.2] (6) at(4.0,0.8){$6$};
		
		\draw [dotted,-] (0.55,-0.5)--(0.55,1);  \node (S00) at(0.55,-0.8){$S_0$};
		\draw [dotted,-] (1.25,-0.5)--(1.25,1);  \node (S2) at(1.15,-0.8){$S_2$};
		\draw [dotted,-] (3.25,-0.5)--(3.25,1); \node (S01) at(3.25,-0.8){$S_0$};
		\draw [dotted,-] (4.45,-0.5)--(4.45,1); \node (S1) at(4.45,-0.8){$S_0$};

			\path[]
				(1) edge node {} (0)
				(2) edge node {} (0)
				(3) edge node {} (1)
			    (5) edge node {} (3)
				(6) edge node {} (3)
				(7) edge node {} (5)
				
				(S00) edge node {$\sigma_3$} (S2)
				(S2) edge node {$\sigma_4$} (S01)
				(S01) edge node {$\sigma_2$} (S1)
				;
				
						
	\end{tikzpicture}
	\caption{
		The convergence due to an honest \SBs followed by a silent period longer than delay bound.
		The $\Delta$-fork based on $w=\cdots HHh0h0h$ with delay $\Delta=1, \theta=4$.
	}
	\label{fig: convergence example}
\end{figure}
\begin{figure}[h]
	\centering
	\begin{tikzpicture}[->,shorten >=1pt,auto,node distance=1cm,-stealth,
		thick,base node/.style={square,draw ,minimum size=48pt}, real node/.style={double,square,draw,minimum size=50pt},scale = 1.7]

	
		\node [draw=\hon, dashed, scale = 2.2] (0)at(0.1,0){$ $};
		\node [draw=\hon, fill=\honB, scale=1.2 ] (1)at(0.8,0.8) {$1$};
		\node [draw=\hon, fill=\honB, scale=1.2] (2)at(1.5,0){$2$};
		\node [draw=\hon, scale=1.2] (3)at(2.3,0.8){$3$};
		\node [draw=\hon, scale=1.2] (5)at(3.5,0){$5$};
		\node [draw=\hon, scale=1.2 ] (8)at(4.8,0.8){$8$};
		
		\path[]
				(2) edge node {} (0)
				(5) edge node {} (2)
				(1) edge node {} (0)
				(3) edge node {} (1)
				(8) edge node {} (3);
		\draw [dotted,-] (0.65,-0.5)--(0.65,1);  \node  (state0) at(0.68,-0.8){$S_0$};
		\draw [dotted,-] (1.35,-0.5)--(1.35,1);  \node  (state1) at(1.35,-0.8){$S_2$};
		\draw [dotted,-] (3.35,-0.5)--(3.35,1);  \node  (state2) at(3.35,-0.8){$S_0$};
		\draw [dotted,-] (4.65,-0.5)--(4.65,1);  \node  (state3) at(4.65,-0.8){$S_0$};
		
		\path[]
				(state0) edge node {$\sigma_3$} (state1)
				(state1) edge node {$\sigma_4$} (state2)
				(state2) edge node {$\sigma_2$} (state3);
	\end{tikzpicture}
	\caption{
		Illustration of failure of the event $S_2\myto{\sigma_4}S_0\myto{\sigma_2}S_0$ in leading to agreement.
		The $\Delta$-fork is based on $w=\cdots HHh0h00h$ with delay $\Delta=2,\theta=4$.
		At a time slot 8, the two fork chains are still competing, 
		even though the system just finishes the transition $S_0 \myto{\sigma_2}S_0$.
		}
	\label{fig:impossible convergence case}
\end{figure}
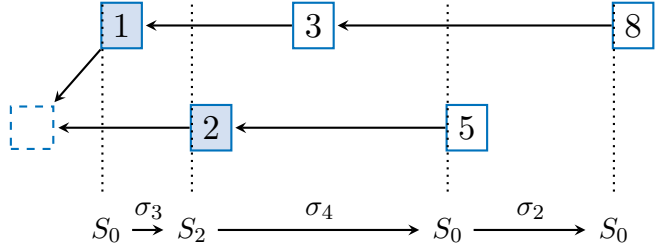

\noindent \textbf{State Transitions and Consistency.}
\label{paragraph: analysis of transitions}
We first provide three scenarios where honest parties have a chance to converge, and what it takes for the adversary to break the convergence.
Our proof will be based on analyzing the probability of these scenarios and the requirement of the adversary.
\Cref{fig: convergence example,fig: high-diff block,fig:impossible convergence case} share the same legends for the types of blocks, and the block with dashed edges on the left end is the one that all honest parties agree on and follow. 
The numbers in each block are the labels defined in \cref{subsec: forks}, i.e., the time slots when these blocks are created.

The first scenario, illustrated in \cref{fig: high-diff block}, shows the convergence due to an honest \BB.
Before time slot 7, there is a competition between honest block 5 and honest block 4 due to the network delay $\Delta =2$.
At time slot 7, the honest \BBs 4, which has been received by all honest parties, beats the \SBs 5 due to its larger weight.
In the semi-Markov chain model, it is the state transition $S_1\myto{\sigma_4}S_0$ that finalizes the competition of the two honest fork chains $1\leftarrow4$ and $2\leftarrow5$.
In this event, to prevent the convergence, the adversary needs to publish blocks with at least $\theta$ weights (e.g., the adversary block $4$ in the graph) before time slot 7. 
Otherwise, all honest parties will follow the heaviest chain $1\leftarrow4\leftarrow7$ at time slot $7$.
So we can conclude that whenever the event $S_1\myto{\sigma_4}S_0$ happens, the adversary needs to publish blocks with at least $\theta$ weights to break the consistency. 
Thus, the frequency of the occurrence of such a pattern will induce a minimum requirement of the adversary mining power in order to break consistency.

The second scenario, illustrated in \cref{fig: convergence example}, shows the convergence due to an honest \SBs followed by a silence period longer than the delay bound.
By time slot 7, if the adversary does not publish a \SBs (e.g., block $6$), 
all honest parties will agree on the chain containing block $5$, because no other blocks compete with the block $5$ in the following $\Delta$ time slots.
Such a situation, characterized by the event $S_0\myto{\sigma_2} S_0$, requires 1 weight from the adversary to break the consistency. 

The third scenario is quite similar to the second one. 
If we replace the \SBs 5 in \cref{fig: convergence example} with an \BB, which is characterized by $S_0\myto{\sigma_4} S_0$ instead of $S_0\myto{\sigma_2} S_0$, all honest parties will agree on the chain containing block $5$. 
The only difference is that it takes the adversary $\theta$ weight to break the consistency. 

Although the above three scenarios suffice for our proof, we provide a carefully constructed scenario, among many scenarios where honest parties cannot reach an agreement, to illustrate the intricacies. 
We can say that the second scenario is characterized by $S_0 \myto{\sigma_2}S_0$. 
However, we cannot say that $S_0 \myto{\sigma_2}S_0$ must fall into the second scenario. 
\cref{fig:impossible convergence case} depicts an example, which can be characterized in the same way as the second scenario ($S_0 \myto{\sigma_3}S_2\myto{\sigma_4} S_0\myto{\sigma_2}S_0$) in the semi-Markov model. 
In this example, honest parties fail to converge even in the absence of the adversary. 
At time slot $2$, two \BB s compete with each other.
The newly mined \SBs $3$ does not finalize the competition since there is no sufficiently long silent period afterward for synchronization. 
Hence, \SB s $5$ and $8$, produced by honest parties, append after different (honest) chains due to network delay.
During the whole period, honest parties cannot reach an agreement on the heaviest chain at any time.
This case shows that unlike the transitions $S_0\myto{\sigma_4}S_0$ and $S_1\myto{\sigma_4}S_0$ (corresponding to the first and the third scenarios), transition $S_0\myto{\sigma_2}S_0$ needs to be analyzed carefully.

\noindent \textbf{Adversary Mining Power Requirement.}
\label{paragraph: alpha}
Combining all the above-discussed state transition scenarios, we can obtain the mining power requirement for adversary parties to break consistency.
It is a minimal requirement since we always consider the worst cases for honest parties and ideal cases for the adversary.
Our analysis depends on counting the \emph{agreement blocks} defined to be those which beat other competing blocks (if any) and thus are accepted by all honest parties (e.g., \BBs 4 in \cref{fig: high-diff block}, \SBs 5 in \cref{fig: convergence example}).

Consider a period with length $L$.
We define $\alpha_L$ as the total weights of the agreement blocks divided by $L$, which is also the mining power requirement for the adversary during this period to break convergence.
Let $\#_L\{E\}$ be the counter of the occurrence of the event $E$ in the given period.
%
As explained in the above scenarios, once the event $S_0\myto{\sigma_2}S_0$ with the previous state not at $S_2$ happens, the adversary needs to produce blocks with at least $1$ weight.
In the events $S_0\myto{\sigma_4}S_0$ or $S_1\myto{\sigma_4}S_0$, the adversary needs to produce blocks with at least $\theta$ weights.
Thus, $\alpha_L$ can be expressed as
\begin{equation*}
	\begin{array}{l}
		\alpha_L = \frac{1}{L}\Big[\#_L(S_0\myto{\sigma_2}S_0) -  \#_L(S_2\myto{\sigma_4}S_0\myto{\sigma_2}S_0 ) 
		  + \theta (\#_L(S_0\myto{\sigma_4}S_0)+\#_L(S_1\myto{\sigma_4}S_0) ) \Big].
		\end{array}	
\end{equation*}
For example, $\alpha_{7}=\frac{\theta}{7}$ in \cref{fig: high-diff block}, because the transition $S_1\myto{\sigma_4}S_0$ requires $\theta$ weights in $L=7$ time slots.
In \cref{fig: convergence example}, $\alpha_{7}=\frac{\theta+1}{7}$, since the transitions $S_0\myto{\sigma_2}S_0$ and $S_0\myto{\sigma_4}S_0$ produce 1 and $\theta$ weight(s), respectively.

It is not tractable to analyze $\alpha_L$ for a fixed $L$ precisely to obtain a closed-form expression. 
Fortunately, using renewal theory and strong law of large numbers, we can show that $\alpha_L$ converges almost surely to a constant $\alpha$, which is the long-run lower bound of minimal required mining power for the adversary. 
The proof of the following lemma is in \cref{appendix: Proof of Lemma 3.2}.

\begin{mylem}
\label{well-defined alpha}
The \alphanames $\alpha := \lim\limits_{L\to\infty}\alpha_L$ is well-defined and it can be calculated as
\begin{align*}
\alpha = \frac{\pi_0 P_{00}\prob{\sigma_2}}{[\prob{\sigma_2}+\prob{\sigma_4}]\sum_{i=0}^2 \pi_i\mu_i}\left( 1- \frac{\pi_2P_{20}}{ \sum_{i=0}^2 \pi_i P_{i0} }      \right)
+\theta \frac{\pi_0P_{00} \prob{\sigma_4} + \pi_1 P_{10}\prob{\sigma_4}   }{ [\prob{\sigma_2}+\prob{\sigma_4}]\sum_{i=0}^2 \pi_i\mu_i },
\end{align*}
where $\pi_i$ and $\mu_i$ are the stationary distribution of the embedded Markov chain and the expected time spent in state $S_i$ before making a transition, respectively.
The exact expression of $\alpha$ is 
\begin{equation}
	\label{eq:alpha}
		\begin{array}{l}
			\alpha = \frac{1}{p_h}\Big[(1-q_H)^\Delta (q_h^2(1-p_h)^{2\Delta} 
			\hspace{5pt} +q_Hq_h(1-p_h)^\Delta(1-q_H)^\Delta+\theta p_hq_H(1-q_H)^\Delta)\Big].
		\end{array}
	\end{equation}
\end{mylem}


We use concentration inequalities to measure the deviation $|\alpha_L - \alpha|$.
Intuitively, the time for transitions between states is a sub-exponential random variable, so the tail probability of the transition time deviating from its expectation has an exponential decay.
Such concentration bounds enable us to establish that over a long time period, the probability that the $\alpha_L$  deviates from its limit will be exponentially small in the deviation size.

\begin{mylem}\label{lem:concentration}
	For a given period $L$ with at least one honest block, the probability of underestimating the \alphaname, namely, 
	$\alpha_L<(1-\delta)\alpha$, is negligible for sufficiently large time $L$.
\end{mylem}

We defer the proof to \cref{appendix: proof of concentration}. 
Due to the dominated convergence theorem, $\e{\alpha_L} \to \alpha$ as $L$ goes to infinity.
The asymptotic result is the best that we can obtain since $\e{\alpha_L}$ differs for different $L$, e.g., for $L<\Delta$, $\alpha_L\equiv0$. 
Although the sub-gaussian property holds for the random variable $\alpha_L$, we can only obtain a sub-exponential tail probability of the deviation $|\alpha_L-\alpha|$ because the time spent on each edge is a sub-exponential random variable.
This lemma shows that we can use $L\alpha$ to approximate the weights of agreement blocks produced by honest parties in the given $L$ slots.


\begin{mythm}[Consistency]
	\label{thm: Consistency}
	The \protocols protocol satisfies the consistency if there exists $\delta >0$ such that
	\begin{equation}
	\label{eq:consistency bound}
		\alpha \geq (1+\delta)\beta,	
	\end{equation}
	where $\beta=q_a + \theta q_A $. 
\end{mythm}
\begin{myproof}
	If there exists a $\delta >0$ satisfying the requirement, there will be a mining power gap between honest parties and the adversary.
	According to \cref{well-defined alpha}, the heaviest honest chain has a higher weight growth rate than the adversary.
    When the leading weights of the heaviest honest chain are $\theta$, 
	then at least one agreement block is produced by honest parties, 
	and the adversary fails to announce blocks with equal weights to break consistency.
	Therefore, the consistency property holds for $k>\frac{\theta}{\alpha-\beta}$ with high probability according to \cref{lem:concentration}.
\end{myproof}

Intuitively, $\beta$ is the maximal rate of block weights that the adversary can produce.
When $\beta>\alpha$, the adversary can break the consistency by announcing blocks with enough weights with high probability whenever the transitions $S_0\myto{\sigma_2}S_0$, $S_0\myto{\sigma_4}S_0$ and $S_1\myto{\sigma_4}S_0$ happen.

%% file: analysis/comparison.tex
\noindent \textbf{Comparison with Nakamoto's Consistency.}
To show the improvement brought by our method on the consistency bound, 
we theoretically compare \alphaname s for the adversary to break consensus in the Nakamoto protocol with and without \protocol.
It suffices to only compare with the original Nakamoto protocol because many PoW-based protocols are derived from it, and the improvement can potentially be carried over to other Nakamoto variants as we will discuss in \cref{sec:literature}.

Since the original Nakamoto protocol can be viewed as a special case of Ironclad with the probability of \BB s set to be 0, 
we can calculate the \alphanames for Nakamoto by plugging $q=0$ into \cref{eq:alpha} to get $\alpha_0=p_h(1-p_h)^{2\Delta}$.
Note that \cite{better_method} provides almost the same consistency bound as $\alpha_0$ by considering the model with a slightly different pattern decomposition.
The bound \cite{better_method}, essentially $\alpha_0$, 
serves as a good benchmark because it provides a precise expression of Nakamoto consistency bound, 
and it is comparable with our result in the same theoretical framework.

We define a relative performance $\mathcal{R}=\frac{\alpha/(\theta q + 1 -q)}{\alpha_0/1}$ to compare the \alphanames (normalized by the mean of block weight) for the adversary to break consensus in Nakamoto consensus with and without \protocol. 
To show the improvement when using \protocol, we will show that $\mathcal{R}$ is larger than 1 for a broad range of parameters.


\begin{mythm}
\label{prop:R>1}
The relative performance ratio $\mathcal{R}> 1$ for all proper parameters:
$$
0<p_h,q<1, \ \Delta \geq 1,\ \theta \geq \max\{\underline\theta,1\} >0,
$$
where the lower bound $\underline\theta$ for the weight of \BB s is defined in \cref{eq:theta lower bound}.
\end{mythm}

We provide a sketch of the proof here.
We first treat $p_h,\Delta,q$ as constants and view $\mathcal{R}$ as a function of $\theta$.
Then, we show that $\mathcal{R}$ is monotonically increasing in $\theta$. 
Thus, it suffices to show $\mathcal{R}>1$ for $\theta=1$ using the weighted power mean inequality.
The full proof with details is in \cref{appendix: proof of R}.

Another way to view the consistency bound of the Nakamoto protocol is from \cite{tighter_bound,Nakamoto-Win} which state that consistency holds under the following condition,
\begin{equation}
\label{eq:tight bound}
    \frac{1}{p_h} +\Delta < \frac{1}{p_a},
\end{equation}
where $p_h$ and $p_a$ are defined in \cref{subsection:model}, representing the honest and adversarial mining rates, respectively.
Note that the fraction of adversarial mining power is $\rho=\frac{p_a}{p_h+p_a}$ and the block production time (in time slots normalized by the delay bound $\Delta$) is $c=\frac{1}{(p_h+p_a)\Delta}$.
Therefore, \eqref{eq:tight bound} can be visualized in \cref{fig:final result} (so do our consistency bound \eqref{eq:consistency bound} and the bound in \cite{better_method}).

The consistency bound of the original Nakamoto protocol \eqref{eq:tight bound} is proven to be tight in \cite{tighter_bound,Nakamoto-Win}, 
which is intuitive since the LHS of \eqref{eq:tight bound} indicates how the network delay affects the honest chain growth rate in the worst case.
If the adversary has a higher mining rate than this ``discounted'' honest mining rate, 
then the system is likely to be insecure since the adversary has enough power to break the agreement among honest parties.

As depicted in \cref{fig:final result}, 
our the consistency bound of Nakamoto+\protocols outperforms the tight bound \eqref{eq:tight bound} for Nakamoto. 
It means that our method \protocol can go beyond the theoretical limit of the original Nakamoto protocol.
The reason is that our method leads to fundamental improvement on the consensus by changing the protocol from ``longest chain rule'' to ``heaviest chain rule''. 

\noindent \textbf{A Lower Bound of $\theta$.}
The example in \cref{fig: high-diff block} shows how \BB s lead to consensus by dominating the competing \SB s (corresponding to $\sigma_4$) due to their larger weight.
This outcome requires that $\theta$ be larger than the number of competing \SB s within the preceding and following $\Delta$ time slots at creating such an \BB\ (corresponding to $\sigma_4$).
Hence a naive lower bound is $2\Delta$, which is too large in practice.
The weight parameter $\theta$, jointly with the \qnames $q$, should be a hyperparameter to allow different choices and optimization. 
We can trade off the value of $\theta$ and the requirement holding probability.
We first construct the upper confidence bound of the number of \SB s in $\sigma_4$. 

\begin{mylem}
    \label{lem: setting theta}
    The probability that the number of \SB s $Y$ in the pattern $\sigma_4$ exceeding $\theta$ decays exponentially as $\theta$ grows. More precisely,
    $$
    \prob{Y\geq \theta} \leq \exp\left(-\frac{q_h\Delta \delta^2}{(1-q_H)(2+\delta)}\right),
    $$
    where $\delta = (\frac{1-q_H}{  q_h\Delta})\theta-1$.
\end{mylem}
See proof in \cref{appendix: proof of low theta}.

To satisfy the requirement with probability $1-\epsilon$, we can leverage \cref{lem: setting theta} to construct an lower confidence bound of $\theta$ as
\begin{equation}
    \label{eq:theta lower bound}
	\underline{\theta} := \frac{2 q_h\Delta}{1-q_H} -  \ln \epsilon 
	+ \sqrt{(\ln \epsilon)^2 - \frac{8 q_h\Delta}{1-q_H}   \ln \epsilon}.
\end{equation}

In \cite{asynchronous_networks,better_method}, $p_h = 10^{-13}$, $\Delta = 10^{13} $ time slots,
which corresponds to the case where the average time between two honest blocks is 10 seconds.
By setting a significance level $\epsilon = 10^{-10}$,
we can get $\theta \geq 51.8$, $\forall q\in(0,1)$, 
which is a mild requirement for $\theta$ compared with $2\Delta$.

%% file: analysis/growth_and_quality.tex
\subsection{Liveness} 
\label{subsec:growth and quality}

As an essential property of a public ledger, \emph{liveness} relies on the fundamental chain properties: \emph{chain quality} and \emph{chain growth}, which we establish in this section. 

Chain growth indicates the growth rate in terms of the weight of the heaviest honest chain. 
Since the heaviest chain may switch as the fork grows, it is necessary to track the weight growth rate of the $\Delta$-fork instead of a fixed single chain.

\begin{mylem}[Chain Growth] 
    \label{lem:chain growth}
    Given $S\in [L]$, $t \in[L-S]$ and a $\Delta$-fork $F_L \in \Dfork{{w}_{1:L},{w'}_{1:L}}$ over a lifetime of $L$ slots, for any $\delta>0$, the weight growth of the fork $\diffFC{t+S}-\diffFC{t}$ is at least $(1-\delta)S p_h\frac{1-q+q\theta}{1+p_h\Delta}$ with probability $1-\exp({-\Omega(S{\delta}^2)})$.
\end{mylem}


The proof is in \cref{appendix:chain growth}, and we provide a brief sketch here.
At time $t$, the honest parties hold chains in $F_t$ with cumulative weights at least $\diff{\overline{{F_{t -{\Delta}}}}}$. 
Therefore, $\diffFC{t+S}-\diffFC{t}$ measures the minimum growth of cumulative weights during $S$ time slots, which is characterized by parameter $g=\frac{p_h(1-q+q\theta)}{1+p_h\Delta}$. 



\begin{mylem} [Chain Quality]
  \label{lem:chain quality}
  For a given period with $L$ time slots, let $F_{t}\in\mathcal{F}({ {w}_{1:t},{w'}_{1:t}})$ with corresponding characteristic string $|w|=|w'|=L$ and $\Delta \leq t\leq L$. 
  For all dominant chains $\chain_{t-\Delta}\subset\overline{{F_{t -{\Delta}   }}} $ such that $\diff{\chain_{t-\Delta}} =\diff{\overline{{F_{t -{\Delta}   }}}} = D$, and all chain segments $\mathcal{S}\subset \mathcal{C}_{t-\Delta}$ such that the first block of $\mathcal{S}$ is discovered by honest parties and the last block is the head of $\mathcal{C}_{t-\Delta}$, the weights contributed by honest blocks in $\mathcal{S}$ is at least $1-(1+\delta)D(\frac{p_a}{p_h}+p_a \Delta)$ with probability $1-\exp({-\Omega(\delta^2D)})$.
\end{mylem}

Chain quality is the fraction of the honest weight in the heaviest honest chain.
Our method Ironclad guarantees the chain quality to be above the fraction 
$f = 1-(1+\delta)(\frac{p_a}{p_h}+p_a \Delta)$.
The proof is deferred to \cref{appendix:chain quality}.



It is now straightforward to establish the liveness property for Nakamoto consensus equipped with \protocol, which is similar to previous works~\cite{tighter_bound,backbone_protocol,asynchronous_networks,ren2019analysis}.
The intuition is that chain growth and chain quality guarantee the occurrence of honest blocks in the heaviest chain and consistency, which we have established in \cref{thm: Consistency}, 
ensures that there exists a time such that honest parties will agree on the same heaviest chain.

\begin{mythm}[Liveness]
Nakamoto's protocol equipped with \protocols satisfies \emph{liveness} if \emph{chain growth, chain quality} and \emph{consistency} hold.
\end{mythm}

%% file: discussion.tex
\section{Selection of Parameters} 
\label{section: Discussion}

\input{discussion/discussion_parameters.tex}

\input{discussion/random_walk.tex}

%% file: discussion/discussion_parameters.tex
\subsection{Optimal $q$}
\label{subsec: selection of q}

The parameter $q$ in Ironclad controls the frequency of \BB s among all valid blocks,
which is flexible by design.
According to \cref{thm: Consistency}, consistency is satisfied for $\beta<\alpha$, or equivalently,  
$$
\frac{p_a}{p_h} < \frac{\alpha}{p_h(1-q+\theta q)}.
$$
To select a reasonable $q$, we define the tolerance ratio of mining power between the adversary and honest parties. 
\begin{mydef}[Tolerance Ratio]
    \label{def: tolerance ratio}
The tolerance ratio $\mathcal{A}$ is defined as
$$
\mathcal{A} = \frac{\alpha}{p_h(1-q+\theta q)}.
$$
\end{mydef}
The analytical expression for the tolerance ratio is shown in \cref{eq:expression of A}.
In order to deal with it mathematically, 
we also derive a close approximation $\tilde{\mathcal{A}}$ in \cref{eq:A tilde}.
We show that $\tilde{\mathcal{A}}$ fits $\mathcal{A}$ well in \cref{fig:adv_ratio in theory}.
We find three interesting properties of $\mathcal{A}$,
which reveals some unique features of our method.
\begin{mythm}
  \label{prop:adversary ratio properties}
  For any $p_h$, $\Delta$ and $\theta\geq \max(\underline\theta,1)$, 
\begin{enumerate}
    \item the maximizer $q^*$ for $\mathcal{A}$ exists in $(0,1)$;
    \item a unique maximizer $q^*(\theta)$ of $\tilde{\mathcal{A}}$ exists in $(0,1)$. Moreover, $q^*(\theta)$ decreases in $\theta$;
    \item the optimal value $\mathcal{A}^*_{\theta}$ of $\mathcal{A}$ for each $\theta$ increases in $\theta$.
\end{enumerate}
\end{mythm}

The intuition of the proof, which is in \cref{appendix: proof of A}, can be explained using the first chart in \cref{fig:adv_ratio in theory}. 
The system consensus benefits (tolerance ratio $\mathcal{A}$ increases) from increasing $q$ when $q$ is small. 
This is because \BB s can effectively reduce the forks caused by \SB s. 
However, when $q$ exceeds a certain threshold $q^*$, $\mathcal{A}$ starts to decrease because the competition among \BB s themselves becomes the main issue as \SB s become less.

The unimodality of tolerance ratio in $q$ provides useful insights. 
If we fix $p_h$, $\Delta$ and $\theta$, there will exist a non-trivial maximizer $q^*$ for it.
Therefore, we can obtain the optimal $q^*$ by maximizing $\mathcal{A}$ for each $\theta$, which can be numerically solved.
The second chart of \cref{fig:adv_ratio in theory} illustrates the corresponding results for specific $p_h$ and $\Delta$ in the Ethereum setting.
We find a decreasing trend for $q^*$ as $\theta$ grows, which conforms to our intuition.
The approximation result for $q^*(\theta)$ also explains the intuition mathematically.
It is obvious that the approximation curve fits the sample points very well.

The third property is depicted in the third chart of \cref{fig:adv_ratio in theory}.
The mining power gap in the original Nakamoto protocol is $p_h-p_a$, 
and it is scaled up to be $(p_h-p_a)(1-q+q\theta)$ in Nakamoto+\protocol\ due to the effect of \BB s, 
which indicates that the new one can tolerate more adversarial mining power. 


\def \green {ForestGreen}
\def \blue {RoyalBlue}
\def \orange {orange}
\def \marksquare {square}
\def \marko {o}
\def \marktri {triangle}

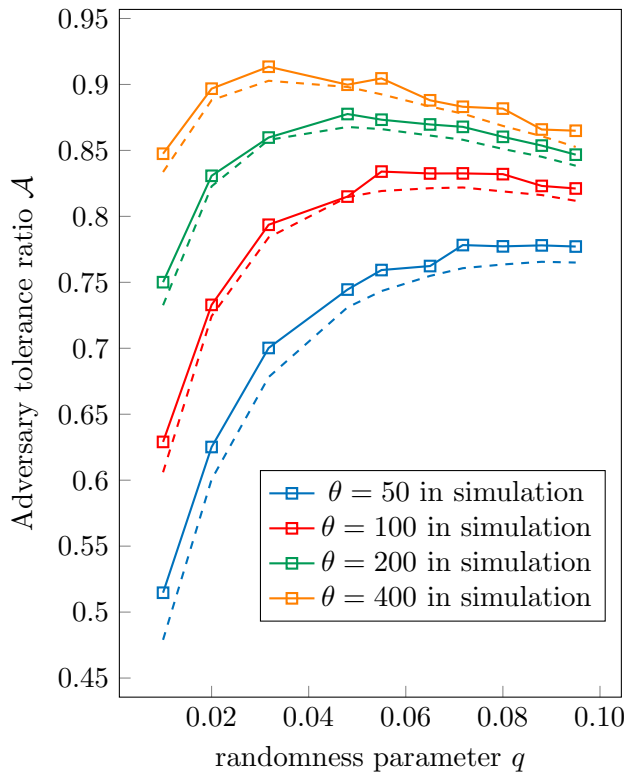
\begin{figure}[hbtp]
    \centering
\begin{tikzpicture}
    \begin{axis}[
        height = 0.65\textwidth,
        width = 0.5\textwidth,
        xlabel = \qname\ $ q $,   
        scaled ticks=false,
        ylabel = Adversary tolerance ratio $\mathcal{A}$,
        xmin=0.001,
        scaled ticks=false,
        xtick pos = left,
        ytick pos = left,
        x tick label style={/pgf/number format/.cd,std,precision=2,
                            /pgf/number format/fixed,     
                            /pgf/number format/fixed zerofill,
                             },
        legend style={at={(0.28,0.22)}, anchor=west, nodes={scale=1, transform shape}}
    ]
    
    \addplot [color=\blue , mark=\marksquare, thick ] table [x index=1,y index=4, col sep = comma] {tex_figure/adv_ratio/theta50.csv};
    \addplot [color=red , mark=\marksquare , thick] table [x index=1,y index=4, col sep = comma] {tex_figure/adv_ratio/theta100.csv};
    \addplot [color=\green , mark=\marksquare , thick] table [x index=1,y index=4, col sep = comma] {tex_figure/adv_ratio/theta200.csv};
    \addplot [color=\orange , mark=\marksquare , thick] table [x index=1,y index=4, col sep = comma] {tex_figure/adv_ratio/theta400.csv};
    
    \addplot [color=\blue, dashed, thick] table [x index=1,y index=5, col sep = comma] {tex_figure/adv_ratio/theta50.csv};
    \addplot [color=red, dashed, thick] table [x index=1,y index=5, col sep = comma] {tex_figure/adv_ratio/theta100.csv};
    \addplot [color=\green, dashed, thick] table [x index=1,y index=5, col sep = comma] {tex_figure/adv_ratio/theta200.csv};
    \addplot [color=\orange, dashed, thick] table [x index=1,y index=5, col sep = comma] {tex_figure/adv_ratio/theta400.csv};
    \legend{
        $\theta = 50$ in simulation,
        $\theta = 100$ in simulation,
        $\theta = 200$ in simulation,
        $\theta = 400$ in simulation,
    }
    \end{axis}
    
\end{tikzpicture}
    \caption{The adversary tolerance ratio $\mathcal{A}$ in theory and simulation with parameter settings: $p_h=\frac{3}{4\Delta}$, $\Delta=10^{13}$. 
    The dashed curves with the same colors correspond to theoretical results.
    The data points are computed by using the Definition \ref{def: tolerance ratio}, where the values of $\alpha$ are obtained from simulation with corresponding $\theta$ and $q$. 
    The simulation settings are explained in detail in \cref{section: Numerical Experiments}.
    }

    \label{fig:adv_ratio: simulation vs theory}
\end{figure}

Furthermore, \cref{fig:adv_ratio: simulation vs theory} compares theoretical $\mathcal A$ with numerical simulation.
We can observe that the properties in \cref{prop:adversary ratio properties} still hold in simulation. 
The optimal \qname\ $q$ with respect to adversarial tolerance ratio $\mathcal{A}$ gradually decreases as $\theta$ ascends. 
It is also observable that $\mathcal{A}_\theta^*$ grows as $\theta$ grows. 
Compared with simulation data, theoretical results serve as relatively precise lower bounds for each case.
In theory, we always consider the worst cases, 
while simulation results are obtained in the sense of average. 
For example, a pattern `hh' is always assumed to cause a fork in theory, 
but it may not cause a fork in reality, 
because the two blocks may be mined by the same miner in consecutive time slots.

%% file: discussion/random_walk.tex
\subsection{Discussion of $\theta$}
\label{subsec: discussion of theta}

The choice of $\theta$ cannot be analyzed by an optimization formulation due to a tradeoff between $\theta$ and confirmation time. 
We build a random walk model that can characterize the attack process in the Nakamoto protocol equipped with \protocols to reveal such a trade-off. 

Suppose that the condition of \cref{thm: Consistency} is satisfied, 
then according to \cref{well-defined alpha}, there exists $\gamma>0$ such that
\begin{equation}
    g = (1+\gamma)p_a (1-q+q\theta),
    \label{eq:margin}
\end{equation}
where $g$ is the chain weight growth rate in \cref{lem:chain growth}.
We analyze the attack process by studying the competition between adversarial chains and the heaviest honest chains. 
Note that the time interval between any two consecutive blocks is a geometric random variable that is ``memoryless" in our slot-based model.
Based on this property, we use a two-dimensional Markov chain to describe the competition.
We use $(X_n,Y_n)$ to denote the state at $n$-th step, representing that the heaviest honest chain leads the adversarial chain by $X_n$ \SB s and $Y_n$ \BB s. 
The initial state is assumed to be $(0,0)$ without loss of generality.
Whenever honest/adversarial parties produce a block, the corresponding coordinate will increase/decrease by $1$.


Denote $\tilde{q}$ the conditional probability for a block to be an \BBs in the honest heaviest chain. 
Clearly, it is larger than $q$.
Intuitively, \BB s can beat competing \SB s and therefore have a better chance to be accepted in the heaviest chain.
The intuition is also supported by simulation data in \cref{section: Numerical Experiments}.
As for the adversary, the conditional probability is still $q$ according to the model assumption.
In general, it is intractable in this model to obtain exact theoretical expressions of $\tilde{q}$ and $\gamma$, so we estimate them by using empirical values in simulations without adversarial attacks.
%
Now we can formulate the state transition probabilities:
$$
(X_{n+1},Y_{n+1} ) = \left \{
\begin{array}{lcl}
(X_{n} + 1, Y_n), \text{ w.p. } \frac{(1+\gamma)(1-\tilde{q})}{2+\gamma} ,\\
(X_{n} - 1, Y_n), \text{ w.p. } \frac{1-q}{2+\gamma},\\
(X_{n}, Y_n + 1), \text{ w.p. } \frac{(1+\gamma)\tilde{q}}{2+\gamma},\\
(X_{n}, Y_n - 1), \text{ w.p. } \frac{q}{2+\gamma}.\\
\end{array}
\right.
$$



We then define the failure rule of attacks, namely, the condition when the adversary gives up the attack.
In \cite{asynchronous_networks,better_method}, the adversary gives up the current attack upon the following situation:
(1) the heaviest chain exceeds the adversarial private chain by at least one block;
(2) the adversary does not mine blocks within the following $\Delta$ time slots;
(3) next block is mined by the honest parties.
In this model, we simplify the failure condition by setting a general threshold:
$$
\tau = \{n: X_n +\theta Y_n \geq \text{threshold} \},
$$
which is the stopping time when the adversary gives up the current attack and starts a new one.
During the random walk, the meaningful message is the distribution of the following statistic:
$$
\text{adversarial fork length} = \sum_{n=1}^\tau \id{X_n<X_{n-1}\text{ or }Y_n<Y_{n-1}}.
$$
A longer adversarial fork length leads to a longer confirmation time in applications such as public ledger.

We resort to the Monte Carlo algorithm to obtain numerical solutions. 
For each attack, the Monte Carlo algorithm samples a path from the initial point to a point beyond the area of the plane $\{(x,y)|x +\theta y < \text{threshold}\}$.
For each instance, $q$ is set to be optimal corresponding with each $\theta$ (see the detailed setting in \cref{section: Numerical Experiments}).
We repeat the random walk for $1\times 10^8$ times with the estimators ${\gamma} = 0.5$ corresponding to the honest mining power $p_h=\frac{1}{\Delta}$ and  adversarial mining power $p_a = \frac{1}{3\Delta}$ (i.e., $\rho = \frac{1}{4}$). 
The original Nakamoto protocol can be treated as a special case when $q=\tilde{q}=0$ in the previous random walk model.
The conditional probability $\tilde{q} = 0.12, 0.084, 0.065,0.063,0.053$ for $\theta = 100,200,300,400,500$, respectively.
We set the threshold to be $2$.
In this way, we can obtain the adversarial fork length distribution, 
thus estimating its mean and the tail probabilities.

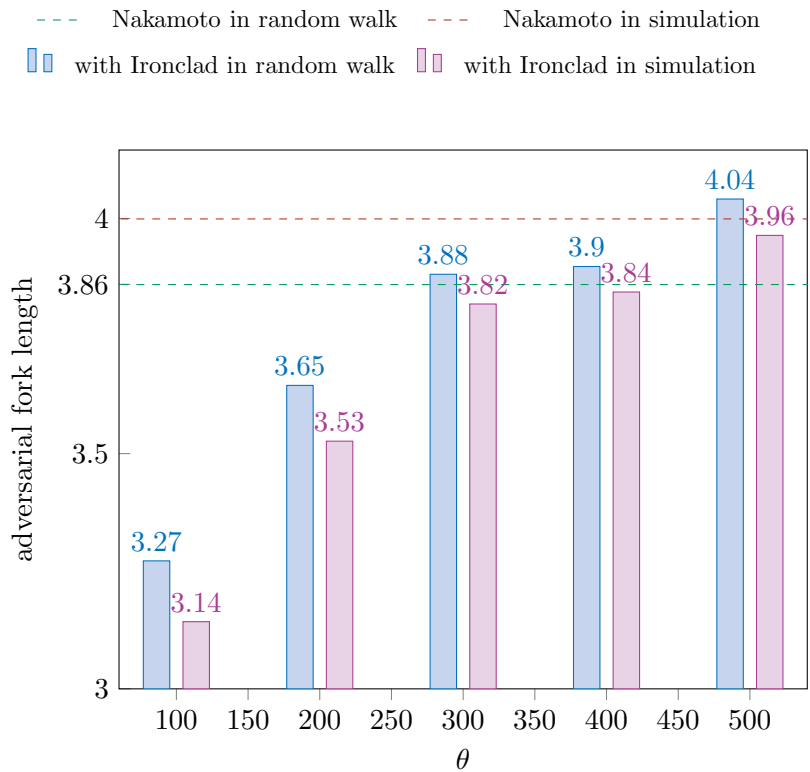
\begin{figure}[hbtp]
    \centering
    \begin{tikzpicture} 
        \begin{customlegend}[legend columns=2,
            legend style={ draw=none,column sep=2ex,nodes={scale=0.85, transform shape}},
            legend entries={ \text{Nakamoto in random walk}, \text{Nakamoto in simulation } }]
            \addlegendimage{color=ForestGreen,dashed }
            \addlegendimage{color=BrickRed,dashed}
        \end{customlegend}   
    \end{tikzpicture} 

    \centering
\begin{tikzpicture}

    \begin{axis}[
        height = 0.53\textwidth,
        width = 0.65\textwidth,
        x tick label style={/pgf/number format/1000 sep={}},
        ylabel = adversarial fork length,
        xlabel = $\theta$,
        ymin = 3,
        ytick ={3,3.5,3.86,4},
        xtick pos = left,
        ytick pos = left,
        legend style={at={(0.4,1.2)},nodes={scale=0.85, transform shape},column sep=1.4ex,draw=none,
        anchor=north,legend columns=-1},
        ybar=5pt,
        grid=minor,
        nodes near coords,
    ]

    \addplot [ color=RoyalBlue, fill = RoyalBlue!20] table [x index=2,y index=1, col sep = comma ] {tex_figure/attacks.csv};
    \addplot [ color=Mulberry, fill=Mulberry!20] table  [x index=2,y index=4, col sep = comma] {tex_figure/attacks.csv};
    \draw [color=ForestGreen, dashed](50,3.85943875)--(550,3.85943875);
    \draw [color=BrickRed, dashed](50,3.9989578)--(550,3.9989578);
    \legend{with \protocol\ in random walk, with \protocol\ in simulation}
    \end{axis}
\end{tikzpicture}
    \caption{This figure depicts the tradeoff between $\theta$ and the expected adversarial fork length.
    The dashed brown and green lines are the expected adversarial fork length of the original Nakamoto in theory and simulation, respectively.
    }
    \label{fig:tradeoff between theta and confirmation time}
\end{figure}

The numerical experiments show that as $\theta$ grows, the expected adversarial fork length will also increase, thus inducing a longer confirmation time. 
\cref{fig:tradeoff between theta and confirmation time} depicts that $\theta$ cannot be too large in order to defend against consensus attacks effectively, since otherwise, it will make the confirmation time too long.
This study also shows that the Nakamoto protocol equipped with \protocols outperforms the original one regarding the expected adversarial fork length in this random walk model and simulation in \cref{section: Numerical Experiments}.

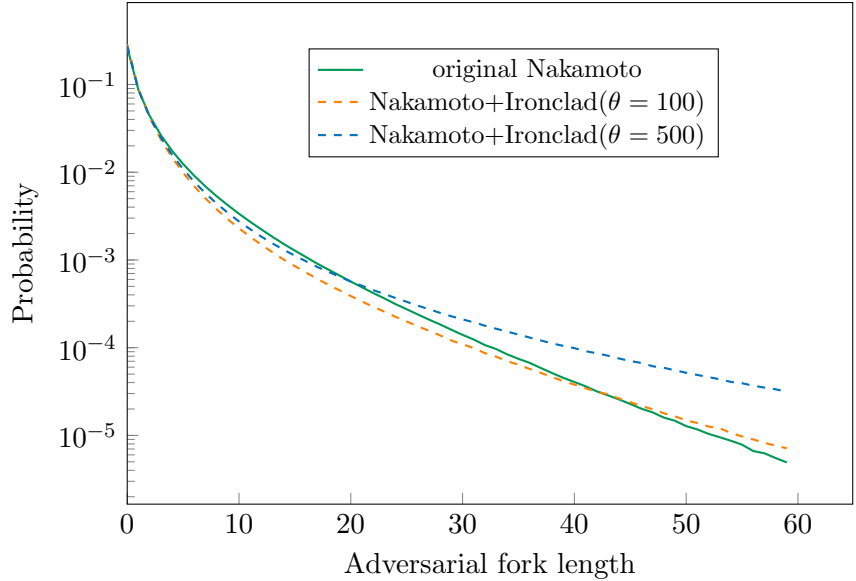
\begin{figure}[hbtp]
    \centering
\begin{tikzpicture}
    \begin{axis}[
        height = 0.5\textwidth,
        width = 0.68\textwidth,
        ymode = log,
        xlabel = Adversarial fork length, 
        scaled ticks=true,
        ylabel = Probability,
        xmin=0,
        xtick pos = left,
        ytick pos = left,
        x tick label style={/pgf/number format/.cd,std,precision=0,
                            /pgf/number format/fixed,     
                            /pgf/number format/fixed zerofill,
                             },
        scaled ticks=true,
        ytick={0.000001,0.00001,0.0001,0.001,0.01,0.1},
        yticklabels ={$10^{-6}$,$10^{-5}$,$10^{-4}$,$10^{-3}$,$10^{-2}$,$10^{-1}$,}, 
        legend style={at={(0.25,0.8)}, anchor=west, nodes={scale=0.9, transform shape}}
    ]
    
   \addplot [color=ForestGreen , thick] table [x index=0,y index=1, col sep = comma] {tex_figure/confirmation_rule.csv};
   \addplot [color=orange , dashed , thick] table [x index=0,y index=2, col sep = comma] {tex_figure/confirmation_rule.csv};
   \addplot [color=RoyalBlue , dashed , thick] table [x index=0,y index=6, col sep = comma] {tex_figure/confirmation_rule.csv};
   
    \legend{
        original Nakamoto,
        Nakamoto+Ironclad($\theta=100$),
        Nakamoto+Ironclad($\theta=500$),
    }
    \end{axis}
    
\end{tikzpicture}
    \caption{This graph depicts the probability for an execution of Nakamoto and the upgraded Nakamoto to sustain a fork of a particular length.}
    \label{fig: confirmation rule}
\end{figure}

\cref{fig: confirmation rule} shows the distributions of adversarial fork length for different weight parameters.
It is observable that a larger $\theta$ also worsens the tail probabilities of adversarial fork length.
The two results together illustrate the trade-off between security and confirmation time:
\emph{a larger $\theta$ leads to a better security threshold but longer confirmation time.}
The intersection points of tail probabilities for the original and upgraded protocols are $(43,2.9\times 10^{-5})$ and $(20,5.67\times 10^{-4})$ for $\theta=100$ and $\theta=500$, respectively.
Indeed, the tail probabilities of \protocol s are worse than that of the original Nakamoto protocols.
However, the required length where the upgraded one becomes worse is quite large and infrequent to confront in practice.

%% file: numerical.tex
\section{Numerical Experiments} 
\label{section: Numerical Experiments}

Our result (\cref{fig:final result}) about adversarial tolerance is based on the analysis of consistency \cref{thm: Consistency}, which shows a significant improvement on the consistency bound by \protocol.
Introducing \BB s  will also affect other properties of blockchain consensus.
This section aims to provide numerical experiments on estimations and comparisons to show the efficacy of \protocol.

\subsection{Setup}

We simulate a fully connected network with 1000 honest miners and network delay bound $\Delta=10^{13}$ time slots.
In the simulation, the adversary has a much lower network delay, maintains its own private chains, and releases its blocks at proper time to prevent the convergence of honest miners as described in \cref{paragraph: analysis of transitions}.

We measure the performance in different adversarial mining power proportions $\rho$ and block generating rates $p$ normalized by the network delay $\Delta$.
In each parameter setting, we keep $\theta=500$ and set corresponding $q$ to be the optimal value using the theoretical results in \cref{subsec: selection of q}.
For example, when we set $\rho=0.25$, $p=0.5$ blocks per $\Delta$ time slots, the optimal $q=0.02475$ for $\theta=500$.

\subsection{Measurement}

\def \crowname {block generating rate $p$}
\def \crowindex {2}
\begin{figure*}[htbp]
    \centering
    \begin{tikzpicture}
        \begin{customlegend}[legend columns=4,legend style={draw=none,column sep=2ex,nodes={scale=0.8, transform shape}},
            legend entries={
                            \text{Nakamoto+Ironclad (by weight)} ,
                            \text{Nakamoto+Ironclad (by number)} ,
                            }]
            \addlegendimage{mark=\markironw,color=\colorironw}
            \addlegendimage{mark=\markironl,color=\colorironl}  
        \end{customlegend}
    \end{tikzpicture}

    \centering
    \begin{tikzpicture}
        \begin{customlegend}[legend columns=4,legend style={draw=none,column sep=2ex,nodes={scale=0.8, transform shape}},
            legend entries={
                            \text{     Nakamoto   },
                            \text{Security threshold of Nakamoto}
                            }]
            \addlegendimage{mark=\marknakal,color=\colornakal}
            \addlegendimage{dashed,color=\colornakal}
        \end{customlegend}
    \end{tikzpicture}

    \advance\leftskip-0.23cm
    \begin{tabular}{rrr}
        \begin{tikzpicture}[scale=0.5]
            \begin{axis}[
                height = 0.3\textwidth,
                width = 0.5\textwidth,
                xlabel = (a)   adversarial mining power proportion $ \rho$,   
                ylabel = Chain quality,
                ymin=0,
                xmin=0,
                ymax=1.03,
                xtick pos = left,
                ytick pos = left,
                legend style={at={(1.1,0.5)}, anchor=west, nodes={scale=1, transform shape}}
            ]

            \addplot [color=\colorironl , mark=\markironl , thick ] table [x index=1,y index=8, col sep = comma] {tex_figure/simulation/single-adv-0.csv};
            \addplot [color=\colorironw , mark=\markironw , thick] table [x index=1,y index=9, col sep = comma] {tex_figure/simulation/single-adv-0.csv};
            \addplot [color=\colornakal , mark=\marknakal , thick] table [x index=1,y index=8, col sep = comma] {tex_figure/simulation/single-adv-1.csv};
            \addplot [color=\colornakal, dashed, thick]table {
                0.382 0
                0.382 1
            };

            \end{axis}
        \end{tikzpicture}
        
        \begin{tikzpicture}[scale=0.5]
            \begin{semilogxaxis}[
                height = 0.3\textwidth,
                width = 0.5\textwidth,
                height = 0.3\textwidth,
                width = 0.5\textwidth,
                log basis x = 2 ,log basis x = 2 ,
                xlabel = (b)   \crowname,   
                ylabel = Chain quality,
                ymin=0,
                xmin=0.01,
                xmax=150,
                ymax=1.03,
                xtick pos = left,
                ytick pos = left,
                xtick={0.01, 0.02, 0.05, 0.1, 0.25, 0.5, 1, 2, 4, 10, 25, 50, 100},
                xticklabels ={$\frac{1}{100}$,$\frac{1}{50}$,$\frac{1}{20}$,$\frac{1}{10}$,$\frac{1}{4}$,$\frac{1}{2}$,1,2,4, 10, 25 ,50, 100},
            ]

            \addplot [color=\colorironl , mark=\markironl , thick ] table [x index=\crowindex,y index=8, col sep = comma] {tex_figure/simulation/single-c-0.csv};
            \addplot [color=\colorironw , mark=\markironw , thick] table [x index=\crowindex,y index=9, col sep = comma] {tex_figure/simulation/single-c-0.csv};
            \addplot [color=\colornakal , mark=\marknakal , thick] table [x index=\crowindex,y index=8, col sep = comma] {tex_figure/simulation/single-c-1.csv};
            \addplot [color=\colornakal, dashed, thick]table {
                2.66 0
                2.66 1
            };
            \end{semilogxaxis}
        \end{tikzpicture}
        \begin{tikzpicture}[scale=0.5]
            \begin{axis}[
                height = 0.3\textwidth,
                width = 0.5\textwidth,
                xlabel = (c) adversarial mining power proportion $ \rho$,   
                ylabel = Quality-growth,
                xmin=0,
                ymin=0,
                xtick pos = left,
                ytick pos = left,
                legend style={at={(1.1,0.5)}, anchor=west, nodes={scale=1, transform shape}}
            ]
            
            \addplot [color=\colorironl, mark=\markironl, thick ] table [x index=1,y index=21, col sep = comma] {tex_figure/simulation/single-adv-0.csv};
            \addplot [color=\colorironw, mark=\markironw, thick] table [x index=1,y index=22, col sep = comma] {tex_figure/simulation/single-adv-0.csv};
            \addplot [color=\colornakal, mark=\marknakal, thick] table [x index=1,y index=21, col sep = comma] {tex_figure/simulation/single-adv-1.csv};
            \addplot [color=\colornakal, dashed, thick]table {
                0.382 0
                0.382 0.8
            };
            \end{axis}
        \end{tikzpicture}


        \begin{tikzpicture}[scale=0.5]
            \begin{semilogxaxis}[
                height = 0.3\textwidth,
                width = 0.5\textwidth,
                log basis x = 2 ,xlabel = (d) \crowname,   
                ylabel = Quality-growth,
                xmin=0.01,
                xmax=150,
                ymin=0,
                xtick pos = left,
                ytick pos = left,
                xtick={0.01, 0.02, 0.05, 0.1, 0.25, 0.5, 1, 2, 4, 10, 25, 50, 100},
                xticklabels ={$\frac{1}{100}$,$\frac{1}{50}$,$\frac{1}{20}$,$\frac{1}{10}$,$\frac{1}{4}$,$\frac{1}{2}$,1,2,4, 10, 25 ,50, 100},
            ]
            
            \addplot [color=\colorironl, mark=\markironl, thick ] table [x index=\crowindex,y index=21, col sep = comma] {tex_figure/simulation/single-c-0.csv};
            \addplot [color=\colorironw, mark=\markironw, thick] table [x index=\crowindex,y index=22, col sep = comma] {tex_figure/simulation/single-c-0.csv};
            \addplot [color=\colornakal, mark=\marknakal, thick] table [x index=\crowindex,y index=21, col sep = comma] {tex_figure/simulation/single-c-1.csv};
            \addplot [color=\colornakal, dashed, thick]table {
                2.66 0
                2.66 20
            };
            \end{semilogxaxis}
        \end{tikzpicture}
    \end{tabular}

    \centering

    \caption{
        Chain quality and Quality growth comparison with varying adversarial proportion and block generating rate in a single chain.
        The values of quality-growth, confirmation time and block generating rates are normalized by the network delay $\Delta$.
        }
    \label{fig:single-chain-1}
\end{figure*}
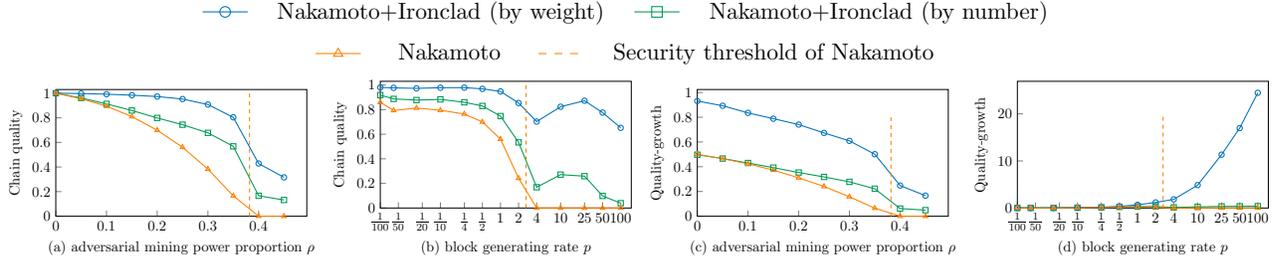

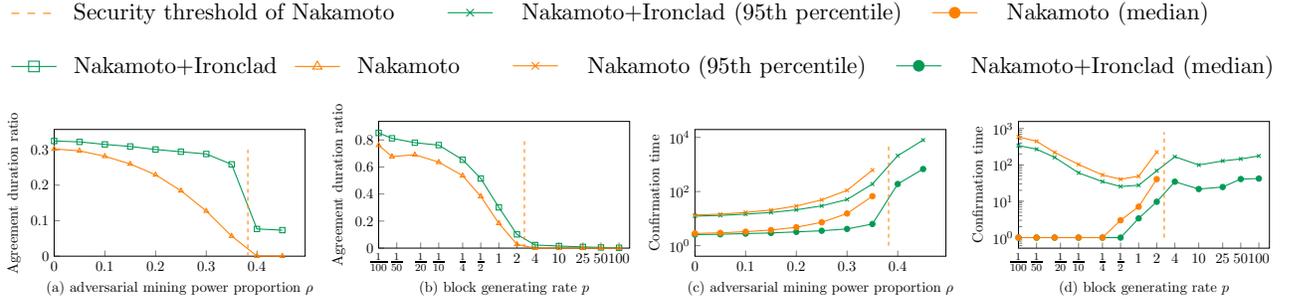
\begin{figure*}
    \centering
    \begin{tabular}{lrr}

        \begin{tikzpicture}
            \begin{customlegend}[legend columns=1,legend style={draw=none,column sep=1ex,nodes={scale=0.8, transform shape}},
                legend entries={
                    \text{Security threshold of Nakamoto}
                                }]
                \addlegendimage{dashed,color=\colornakal}
            \end{customlegend}
        \end{tikzpicture}
         & 
        \begin{tikzpicture}
            \begin{customlegend}[legend columns=4,legend style={draw=none,column sep=2ex,nodes={scale=0.8, transform shape}},

                legend entries={
                    \text{Nakamoto+Ironclad (95th percentile)} ,
                    \text{Nakamoto (median)},
                                }]
                \addlegendimage{mark=\marknine,color=\colorironl} 
                \addlegendimage{mark=\markfive,color=\colornakal}
                
            \end{customlegend}
        \end{tikzpicture}
        
    \end{tabular}

    \begin{tabular}{lrr}
        \begin{tikzpicture}
            \begin{customlegend}[legend columns=4,legend style={draw=none,column sep=1ex,nodes={scale=0.8, transform shape}},
                legend entries={
                                \text{Nakamoto+Ironclad} ,
                                \text{Nakamoto} ,
                                }]

                \addlegendimage{mark=\markironl,color=\colorironl}
                \addlegendimage{mark=\marknakal,color=\colornakal}  
            \end{customlegend}
        \end{tikzpicture}
         & 
         \begin{tikzpicture}
            \begin{customlegend}[legend columns=4,legend style={draw=none,column sep=2ex,nodes={scale=0.8, transform shape}},
                legend entries={
                    \text{Nakamoto (95th percentile)},
                    \text{Nakamoto+Ironclad (median)} ,
                    }]

    \addlegendimage{mark=\marknine,color=\colornakal} 
    \addlegendimage{mark=\markfive,color=\colorironl}
    
            \end{customlegend}
        \end{tikzpicture}
        
    \end{tabular}
    
    \begin{tabular}{rrr}
        \begin{tikzpicture}[scale=0.5]
            \begin{axis}[
                height = 0.3\textwidth,
                width = 0.5\textwidth,
                xlabel = (a) adversarial mining power proportion $ \rho$,   
                ylabel = Agreement duration ratio,
                ymin=0,
                xmin=0,
                xtick pos = left,
                ytick pos = left,
            ]
            
            \addplot [color=\colorironl, mark=\markironl, thick ] table [x index=1,y index=13, col sep = comma] {tex_figure/simulation/single-adv-0.csv};
            \addplot [color=\colornakal, mark=\marknakal, thick] table [x index=1,y index=13, col sep = comma] {tex_figure/simulation/single-adv-1.csv};
            \addplot [color=\colornakal, dashed, thick]table {
                0.382 0
                0.382 0.3
            };
            \end{axis}
        \end{tikzpicture}
        \begin{tikzpicture}[scale=0.5]
            \begin{semilogxaxis}[
                height = 0.3\textwidth,
                width = 0.5\textwidth,
                log basis x = 2 ,xlabel = (b) \crowname,   
                ylabel = Agreement duration ratio,
                ymin=0,
                xmin=0.01,
                xmax=150,
                xtick pos = left,
                ytick pos = left,
                xtick={0.01, 0.02, 0.05, 0.1, 0.25, 0.5, 1, 2, 4, 10, 25, 50, 100},
                xticklabels ={$\frac{1}{100}$,$\frac{1}{50}$,$\frac{1}{20}$,$\frac{1}{10}$,$\frac{1}{4}$,$\frac{1}{2}$,1,2,4, 10, 25 ,50, 100},
            ]
            
            \addplot [color=\colorironl, mark=\markironl, thick ] table [x index=\crowindex,y index=13, col sep = comma] {tex_figure/simulation/single-c-0.csv};
            \addplot [color=\colornakal, mark=\marknakal, thick] table [x index=\crowindex,y index=13, col sep = comma] {tex_figure/simulation/single-c-1.csv};
            \addplot [color=\colornakal, dashed, thick]table {
                2.66 0
                2.66 0.8
            };
            \end{semilogxaxis}
        \end{tikzpicture}
        \begin{tikzpicture}[scale=0.5]
            \begin{axis}[
                ymode=log,
                height = 0.3\textwidth,
                width = 0.5\textwidth,
                xlabel = (c) adversarial mining power proportion $ \rho$,   
                ylabel = Confirmation time,
                xmin=0,
                xtick pos = left,
                ytick pos = left,
            ]
            
            \addplot [color=\colorironl, mark=\markfive, thick ] table [x index=1,y index=7, col sep = comma] {tex_figure/simulation/single-adv-confirm-0.csv};
            \addplot [color=\colornakal, mark=\markfive, thick] table [x index=1,y index=7, col sep = comma] {tex_figure/simulation/single-adv-confirm-1.csv};
            \addplot [color=\colorironl, mark=\marknine, thick ] table [x index=1,y index=12, col sep = comma] {tex_figure/simulation/single-adv-confirm-0.csv};
            \addplot [color=\colornakal, mark=\marknine, thick] table [x index=1,y index=12, col sep = comma] {tex_figure/simulation/single-adv-confirm-1.csv};
            \addplot [color=\colornakal, dashed,thick]table {
                0.382 1
                0.382 5000
            };
            \end{axis}
        \end{tikzpicture}
    
    \begin{tikzpicture}[scale=0.5]
        \begin{semilogxaxis}[
            height = 0.3\textwidth,
            width = 0.5\textwidth,
            log basis x = 2 ,xlabel = (d) \crowname,
            ymode=log,   
            ylabel = Confirmation time,
            xmin=0.01,
            xmax=150,
            ymin=0,
            xtick pos = left,
            ytick pos = left,
            xtick={0.01, 0.02, 0.05, 0.1, 0.25, 0.5, 1, 2, 4, 10, 25, 50, 100},
            xticklabels ={$\frac{1}{100}$,$\frac{1}{50}$,$\frac{1}{20}$,$\frac{1}{10}$,$\frac{1}{4}$,$\frac{1}{2}$,1,2,4, 10, 25 ,50, 100},
        ]
        
        \addplot [color=\colorironl, mark=\markfive, thick ] table [x index=\crowindex,y index=7, col sep = comma] {tex_figure/simulation/single-c-confirm-0.csv};
        \addplot [color=\colornakal, mark=\markfive, thick] table [x index=\crowindex,y index=7, col sep = comma] {tex_figure/simulation/single-c-confirm-1.csv};
        \addplot [color=\colorironl, mark=\marknine, thick ] table [x index=\crowindex,y index=12, col sep = comma] {tex_figure/simulation/single-c-confirm-0.csv};
        \addplot [color=\colornakal, mark=\marknine, thick] table [x index=\crowindex,y index=12, col sep = comma] {tex_figure/simulation/single-c-confirm-1.csv};
        \addplot [color=\colornakal, dashed, thick]table {
            2.66 1
            2.66 800
        };
        \end{semilogxaxis}
    \end{tikzpicture}
    \end{tabular}
    \caption{
        Agreement duration ratio and confirmation time comparison with varying adversarial proportion and block generating rate in a single chain.
        The unbounded values of Nakamoto consensus in (c) and (d) are omitted.
        }
    \label{fig:single-chain-2}
\end{figure*}

In the numerical experiments, we focus on the following four key performance metrics. 

\begin{itemize}
    \item \textbf{Chain quality.}
    We calculate the proportions of honest blocks in the heaviest chain.
    A larger proportion means that honest miners can obtain more rewards (e.g., coinbase and transaction fee, determined by the application level setting of the system).
    We also calculate the honest proportion by weight, since \BB s may be assigned more rewards due to their large weights and higher difficulty in applications. 
    (see \cref{fig:single-chain-1})
    \item \textbf{Quality growth.}
    We calculate the growth rate of honest blocks in the heaviest chain, both by number and by weight.
    For comparison, we normalize the weight growth rate by the expected block weight which varies according to the parameters $q$ and $\theta$.
    (see \cref{fig:single-chain-1})
    \item \textbf{Agreement duration ratio.}
    When all honest parties agree on the heaviest chain due to an agreement block, such agreement will last a while until new forks occurs.
    We calculate the proportion of such periods during the total running time of the execution.
    (see \cref{fig:single-chain-2})
    \item \textbf{Confirmation time.}
    We count the time for each block from production to confirmation by all honest parties. We report the median and 95\% percentile of the confirmation time for all blocks in the simulation horizon. 
    (see \cref{fig:single-chain-2})
\end{itemize}

\subsection{Single Chain: Nakamoto v.s.\ Nakamoto+Ironclad}

The theoretical results established in \cref{subsec: consistency} show that more adversarial mining power is required to break convergence when Nakamoto consensus is equipped with our method.
We conduct comparisons in a wide range of $\rho$ and $p$ to show the efficacy. 

Note that there exists a security threshold of Nakamoto consensus (see \cref{fig:final result}) that indicates the maximum adversarial tolerance and block generating rate. 
When adversarial mining power proportion $\rho$ is too large or the block generating rate $p$ is too high, honest parties cannot reach an agreement with overwhelming probability.
In our experiments of Nakamoto consensus, the performance metrics become either $0$ (e.g., chain quality) or unbounded (e.g., confirmation time) when the parameters violate the security threshold.
We plot a dashed line in each figure to show this security threshold and omit the unbounded values.
Note that Nakamoto+\protocols has a much larger consistency region in terms of $\rho$ and $p$. 

\noindent \textbf{Varying adversarial mining power proportion.}
In this group, we fix $g=1 \text{ block }$ per $ \Delta$  time slots and vary the adversarial mining power fraction ($\rho=0,0.05,0.1,\dots,0.45$).
As shown in \cref{fig:single-chain-1,fig:single-chain-2}, overall, the performance of Nakamoto+\protocols outperforms the original Nakamoto,especially when $\rho$ is large. 
1) \protocols can help tolerate more adversarial mining power. 
The security threshold (dashed line in \cref{fig:single-chain-1,fig:single-chain-2}) indicates that the original Nakamoto can only tolerate at most 0.382 adversarial fraction in this block rate, while the consistency still holds even for $\rho=0.45$ with our method, and our theoretical bound is 0.48 (see \cref{fig:final result} when $c=1$).
2) \protocols can help to reduce the confirmation time. 
The general median and conservative 95th percentile values both illustrate this fact. 
3) \protocols can achieve more robust chain performance when adversarial mining power is not large enough.
We can find that the slopes of curves of the upgraded chain are smaller than those of the original chain when $\rho<0.35$.

Note that there is a gap between the metrics (chain quality and quality-growth) measured by number anb by weights in \cref{fig:single-chain-1} for Nakamoto+\protocol. 
It indicates that the frequency of \BB s is larger than the \qnames $q$ in the heaviest chain, which shows the effect of \BB s in beating competing \SB s and finalizing current forks.

\noindent \textbf{Varying block generating rate.}
In this group, we fix $\rho=0.25$ and $p$ varies from $0.01$ to $100$ blocks per $\Delta$ time slots. 
As shown in \cref{fig:single-chain-1,fig:single-chain-2}, the Nakamoto equipped with \protocols still outperforms the original Nakamoto protocol.
The consensus properties become worse as $p$ increases for the original Nakamoto consensus.
Interestingly, the effect of increasing $p$ on the performance of the Nakamoto+\protocols is not monotone, which is caused by the generating rate of \BB s.
The intuition is that the rate of \BB s increases as $p$ increases, although the corresponding optimal $q$ decreases.
When $p$ is small, the effect of \BB s is not significant due to its low frequency.
As $p$ increases, the honest \BB s are more frequently accepted by all parties, which forces the adversary to give up attacks since honest parties can generate more \BB s in expectation.
When $p$ is very large, the competition among honest \BB s increases, allowing the adversary to hold a heavier private chain. 
Therefore, the performance of chain quality and confirmation time become worse.

%% file: literature.tex
\section{Related Works and Potential Extension} 
\label{sec:literature}

The essence of \protocols is to utilize randomness to enhance the desired properties.
The focus of this paper is to apply it to the original Nakamoto protocol and demonstrate how our method can lead to better consensus performances.
Based on the original idea of Nakamoto consensus, 
there have been many PoW protocols leveraging different innovative ideas to achieve better performances for various applications. 
Although they are pretty complicated and may depend on specific applications, 
we briefly discuss some of these protocols and the potential combination with our method in this section.

\noindent \textbf{Decoupled consensus.}
Many protocols propose different types of blocks to perform different tasks.
BitcoinNG~\cite{BitcoinNG} proposes key blocks (for leader election) and microblocks (for carrying transactions) to improve throughput without comprising security.
FruitChains~\cite{FruitChain} introduces blocks (for leader election) and fruits (for carrying transactions) to decrease the variance of mining rewards and significantly reduce the need for mining pools.
Prism~\cite{Prism} proposes voter blocks (for leader election), transaction blocks (for carrying transactions) and proposer blocks (for packing transaction blocks) to scale up the throughput to approach the physical limit.
Our method differs from these protocols because \BB s are designed to resolve forks using their heavier weight to improve consistency bound.  
This is independent of decoupling functions for leader election and carrying transactions.
Therefore, it is possible to apply \protocols (with adaptation regarding implementation details) to these protocols based on the decoupling idea to achieve better performance.

\noindent \textbf{High forking protocols.}
Simply increasing the block rate will cause more frequent forks and sacrifice security.
GHOST~\cite{ghost} replaces the longest chain rule with the heaviest subtree rule to tolerate forks.
Miners keep track of a tree of blocks instead of a chain and append blocks to the heaviest subtree that is measured by the total \emph{number} of blocks in it.
A series of protocols \cite{inclusive,conflux,EPIC,phantom,SPECTRE} increase reference links and propose their different rules to solve forks in the structure of the directed acyclic graph (DAG).
However, GHOST and Conflux~\cite{conflux} is vulnerable to the balance attack~\cite{balance-attack} in high block rate.
To resist the balance attack, we can apply \protocols to GHOST by adaptively replacing the original \emph{number} of blocks with the \emph{weights} of blocks in a subtree.
Notice that the four patterns in definition~\ref{def:pattern decomposition} play similar roles in analyzing the consistency of GHOST.
Once pattern $\sigma_2$ or $\sigma_4$ occurs, honest parties have chances to figure out the heaviest subtree at the end of the pattern.
If pattern $\sigma_1$ or $\sigma_3$ occurs, it will be impossible for honest parties to identify the heaviest.
Hence, following the same analysis, we can extend \cref{thm: Consistency} to GHOST protocols: GHOST equipped with \protocol\ satisfies the consistency if conditions in \cref{thm: Consistency} are satisfied.

Another recent work GHAST~\cite{ghast} slows down the mining rate when detecting a divergence of computing power.
By setting a low difficulty target $\frac{2^\kappa}{\eta_w}$, GHAST assigns $\eta_w$ weight to blocks achieving this target and 0 for other blocks. 
This is quite similar to the idea of decoupled consensus since blocks with $0$ weight can only carry transactions when switching to this conservative strategy.
Ironclad, in contrast, can avoid the complex switching strategy by following a simple and effective assignment rule. 
Furthermore, our analysis in \cref{subsec: selection of q} shows that the \qnames $q$ is not necessarily the reciprocal of the weight parameter $\theta$.
Our work provides general principles of parameter selection based on an optimization problem to maximize tolerance to the adversary and evaluating the trade-off between security and confirmation latency.

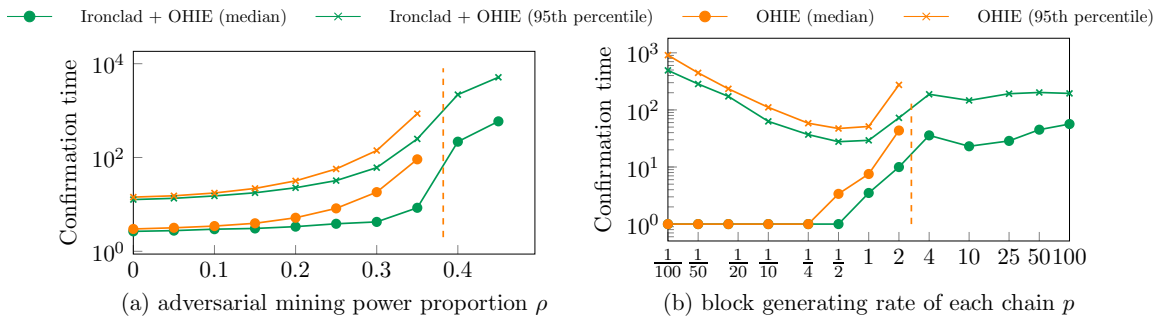
\begin{figure}[htbp]
    \def \crowname {block generating rate of each chain $p$}
    \def \crowindex {2}
    \def \advrowname {adversarial mining power proportion $\rho$}
    \def \advrowindex {1}
    \def \xtick {0.01, 0.02, 0.05, 0.1, 0.25, 0.5, 1, 2, 4, 10, 25, 50, 100}
    \centering
    \begin{tikzpicture}
        \begin{customlegend}[legend columns=4,legend style={draw=none,column sep=2ex,nodes={scale=0.6, transform shape}},
            legend entries={
                            \text{Ironclad + OHIE (median)} ,
                            \text{Ironclad + OHIE (95th percentile)},
                            \text{OHIE (median)},
                            \text{OHIE (95th percentile)},
                            }]
   
            \addlegendimage{mark=\markfive,color=\colorironl}
            \addlegendimage{mark=\marknine,color=\colorironl}  
            \addlegendimage{mark=\markfive,color=\colornakal}
            \addlegendimage{mark=\marknine,color=\colornakal}
        \end{customlegend}
    \end{tikzpicture}

    \advance\leftskip-0.23cm
    \begin{tabular}{ll}

        \begin{tikzpicture}[scale=0.8]
            \begin{axis}[
                ymode=log,
                height = 0.3\textwidth,
                width = 0.5\textwidth,
                xlabel = (a) \advrowname,   
                ylabel = Confirmation time,
                xmin=0,
                xtick pos = left,
                ytick pos = left,
                legend style={at={(1.1,0.5)}, anchor=west, nodes={scale=1, transform shape}}
            ]
            
            \addplot [color=\colorironl, mark=\markfive, thick ] table [x index=\advrowindex,y index=7, col sep = comma] {tex_figure/simulation/parallel-adv-confirm-0.csv};
            \addplot [color=\colornakal, mark=\markfive, thick] table [x index=\advrowindex,y index=7, col sep = comma] {tex_figure/simulation/parallel-adv-confirm-1.csv};
            \addplot [color=\colorironl, mark=\marknine, thick ] table [x index=\advrowindex,y index=12, col sep = comma] {tex_figure/simulation/parallel-adv-confirm-0.csv};
            \addplot [color=\colornakal, mark=\marknine, thick] table [x index=\advrowindex,y index=12, col sep = comma] {tex_figure/simulation/parallel-adv-confirm-1.csv};

            \addplot [color=\colornakal, dashed, thick]table {
                0.382 2
                0.382 8000
            };
            \end{axis}
        \end{tikzpicture}
        
        &
        \begin{tikzpicture}[scale=0.8]
            \begin{semilogxaxis}[
                height = 0.3\textwidth,
                width = 0.5\textwidth,
                ymode=log,
                log basis x = 2 ,xlabel = (b) \crowname,   
                ylabel = Confirmation time,
                xmin=0.01,
                xmax=100,
                ymin=0,
                xtick pos = left,
                ytick pos = left,
                xtick={0.01, 0.02, 0.05, 0.1, 0.25, 0.5, 1, 2, 4, 10, 25, 50, 100},
                xticklabels ={$\frac{1}{100}$,$\frac{1}{50}$,$\frac{1}{20}$,$\frac{1}{10}$,$\frac{1}{4}$,$\frac{1}{2}$,1,2,4, 10, 25 ,50, 100},
                legend style={at={(1.1,0.5)}, anchor=west, nodes={scale=1, transform shape}}
            ]
            
            \addplot [color=\colorironl, mark=\markfive, thick ] table [x index=\crowindex,y index=7, col sep = comma] {tex_figure/simulation/parallel-c-confirm-0.csv};
            \addplot [color=\colornakal, mark=\markfive, thick] table [x index=\crowindex,y index=7, col sep = comma] {tex_figure/simulation/parallel-c-confirm-1.csv};
            \addplot [color=\colorironl, mark=\marknine, thick ] table [x index=\crowindex,y index=12, col sep = comma] {tex_figure/simulation/parallel-c-confirm-0.csv};
            \addplot [color=\colornakal, mark=\marknine, thick] table [x index=\crowindex,y index=12, col sep = comma] {tex_figure/simulation/parallel-c-confirm-1.csv};            \addplot [color=\colornakal, dashed, thick]table {
                2.66 1
                2.66 150
            };
            \end{semilogxaxis}
        \end{tikzpicture}

    \end{tabular}
    \caption{
        Confirmation time comparison with varying block generating rate in each parallel chain and adversarial mining power proportion in 10 parallel chains.
        The dashed lines correspond to Nakamoto security thresholds that are still effective in OHIE since each parallel instance is a Nakamoto chain.
        }
    \label{fig:parallel-chains}
\end{figure}

\noindent \textbf{Parallel chains.}
Parallel chains~\cite{Chainweb,OHIE,Prism}, as special cases of DAG, 
run parallel Nakamoto instances, leading to elegant and provable solutions in scaling blockchain.
These protocols can apply \protocols to improve consensus, 
especially when parallel chains have weak correlations.
As an example, we apply \protocols to OHIE~\cite{OHIE} with ten parallel chains in the same parameter settings ($\rho,p,\theta,q,\Delta$) as the single chain case.
In OHIE, a block is assigned to each parallel chain with equal probability.
Therefore, the balance attack~\cite{balance-attack} is ineffective in OHIE since the adversary cannot concentrate its mining power on attacking a single chain, which allows us to assume that the adversary always chooses honest trailing and behaves similarly to a single chain case.
We omit some metrics such as quality-growth because they can be regarded as the average of all single chains. 
We study the confirmation time since constructing a global ledger relies on a shared confirmation threshold in OHIE, which means that the number of parallel chains affects the system confirmation time. 
As shown in \cref{fig:parallel-chains}, the confirmation time of the parallel chains equipped with \protocols is less than that of original parallel chains overall. 
Compared with single-chain cases in \cref{fig:single-chain-2}c and \cref{fig:single-chain-2}d, parallel chains have higher but similar curves, which shows the comparable efficacy of \protocols improving performance in parallel chains.

\noindent \textbf{Hybrid blockchain-BFT consensus.}
Some works adopt the classical ideas of Byzantine fault tolerant (BFT) consensus to develop PoW-based blockchain protocols~\cite{solida,decker2016bitcoin,kogias2016enhancing,pass2017hybrid}. 
These protocols rely on the establishment of committees to achieve lower latency than Nakamoto-based protocols.
However, the adversary fraction of committees are required to be less than $\frac{1}{3}$ and replenishing committees leads to additional latency. 
A hybrid consensus protocol Thunderella~\cite{thunderella} proposes mechanisms to combine committees and the Nakamoto chain, which allows the system to switch between optimistic and worst-case conditions to achieve low latency and large adversary tolerance.
However, even in the optimal conditions with $\frac{3}{4}$ honest and online committees, a dishonest leader will make the system enter the slow mode with the same latency as Nakamoto consensus.
Our method can still be applied to the Nakamoto part in Thunderella to enhance the ability to resist the adversary when the system is in the slow mode for the same reason as previously discussed.

%% file: appendix.tex






\subsection{Proofs of results in \cref{section: Analysis}}

\subsubsection{Computation of Pattern Length}
The following are four types of patterns defined in \cref{subsec: consistency}.
\begin{enumerate}
  \item $\sigma_1: \{h\}||\{0\}^k, k< \Delta$,
  \item $\sigma_2: \{h\}||\{0\}^k, k\geq \Delta$,
  \item $\sigma_3: \{H\}||\{0,h\}^k, k\leq \Delta$,
  \item $\sigma_4: \{H\}||\{0,h\}^\Delta || \{0\}^*$.
\end{enumerate}
Let $Y$ and $Z$ be the geometric random variables with mean $\frac{1}{p_h}$ and $\frac{1}{q_H}$, respectively. 
\begin{itemize}

    \item For $\e{\sigma_1}$:
    
    The expected number of symbol `0' in $\sigma_1$ is $\e{Y-1|Y<\Delta}$ so the expected length is $\e{Y|Y<\Delta}$. 
    As a result of total expectation formula,
    $$
    \e{Y} = \e{Y|Y<\Delta} \prob{Y<\Delta} + \e{Y|Y\geq\Delta} \prob{Y\geq\Delta}.
    $$
    By the memerylessness property of geometric distribution, 
    $$
    \e{Y|Y\geq\Delta}=\Delta+\e{Y|Y\geq 0 } =\Delta+\frac{1}{p_h},
    $$
    and 
    $$
    \prob{Y\geq\Delta} = \sum_{k=\Delta}^\infty (1-p_h)^kp_h = (1-p_h)^\Delta.
    $$
    Solving for the above equation, we get
    $$
    \e{\sigma_1}=\e{Y|Y<\Delta} =\frac{1}{p_h}-\frac{\Delta (1-p_h)^
\Delta}{1-(1-p_h)^\Delta}.
    $$
    
    \item For $\e{\sigma_2}$:
    
    Similarly, apply the memorylessness property:
    $$
    \e{\sigma_2} = \e{Y|Y\geq\Delta} = \Delta+\e{Y|Y\geq 0 } =\frac{1}{p_h}+\Delta.
    $$
    
    \item For $\e{\sigma_3}$:
    
    The expected number of symbol `0' in $\sigma_3$ is $\e{Z-1|Z<\Delta}$ so similiarly as $\e{\sigma_1}$,
    $$
    \e{\sigma_3}=\e{Z|Z<\Delta} = \frac{1}{q_H}-\frac{\Delta(1-q_H)^\Delta}{1-(1-q_H)^\Delta}.
    $$
    
    \item For $\e{\sigma_4}$:
    
    The computation process is the same as $\e{\sigma_2}$.

\end{itemize}

\subsection{Analysis of the Semi-Markov chain}
\subsubsection{The Expressions for Expected Edge Length}
$$
\begin{array}{lll}
   l_{00} = \frac{\prob{\sigma_2} \e{|\sigma_2|} + \prob{\sigma_4} \e{|\sigma_4|}}{\prob{\sigma_2}+\prob{\sigma_4}}
 & l_{01} = \e{|\sigma_1|}    
 & l_{02} = \e{|\sigma_3|}    \\
   l_{10} = \frac{\prob{\sigma_2} \e{|\sigma_2|} + \prob{\sigma_4} \e{|\sigma_4|}}{\prob{\sigma_2}+\prob{\sigma_4}}
 & l_{11} = \e{|\sigma_1|}  
 & l_{12} = \e{|\sigma_3|}  \\
   l_{20} = \e{|\sigma_4|}  
 & l_{21} = 0   
 & l_{22} = \e{|\sigma_3|}  \\  
\end{array}
$$

\subsubsection{Stationary Distribution of the Embedded Markov chain}
$$
\begin{array}{lcl}
\pi_0 =  \frac{1}{q_H + q_h(1-q_H)^\Delta}(q_H(1-q_H)^\Delta + q_h(1-q_H)^\Delta(1-p_h)^\Delta)\\
\pi_1 =  \frac{1}{q_H + q_h(1-q_H)^\Delta}(q_h(1-q_H)^\Delta (1-(1-p_h)^\Delta))  \\
\pi_2 =  \frac{1}{q_H + q_h(1-q_H)^\Delta}(q_H(1-(1-q_H)^\Delta)) \\  
\end{array}
$$

\subsubsection{Proof of Lemma \ref{well-defined alpha}}\hfill
\label{appendix: Proof of Lemma 3.2}
\begin{myproof}
	We deal with each term in the expression in turn. 
  For simplification, 
	it suffices to consider the first part, and other terms can be computed similarly.
	$$
	\frac{\#_L(S_0\myto{\sigma_2}S_0)}{L} = \frac{\#_L S_0}{L} \cdot \frac{\#_L e_{00} }{\#_L S_0} \cdot \frac{\#_L(S_0\myto{\sigma_2} S_0)  }{\#_L e_{00}}.
	$$
	We compute the limit for each factor. 
	First, we can regard each visit to the state $S_0$ as a renewal. 
	It is because the intervals between each renewal are i.i.d. random variables 
	that only depend on the states of the Markov chain. 
	We use $\mu_{ii}$ to denote the expected transition time between two consecutive visits of $S_i$, then the renewal theorem yields 
	$$
	\frac{\#_L S_0}{L} \myto{a.s.} \frac{1}{\mu_{00}}.
	$$

	$\frac{\#_L e_{00} }{\#_L S_0}$ is the frequency of visiting edge $e_{00}$. 
	Since the embedded Markov chain is ergodic, the state $S_0$ will be visited infinite times. Apply strong law of large numbers(SLLN):
	$$
	\frac{\#_L e_{00} }{\#_L S_0}\myto{a.s.} P_{00}.
	$$
	For the third expression, similarly, $\frac{\#_L(S_0\myto{\sigma_2} S_0)  }{\#_L e_{00}}$ 
	converges to the probability that condition on a visit of the edge $e_{00}$, the visit is caused by $\sigma_2$. 
	The patterns are drawn independently, so the value is
	$$
	\frac{\prob{\sigma_2}}{\prob{\sigma_2}+\prob{\sigma_4}}.
	$$
	
	Therefore, 
	$$
	\frac{\#_L(S_0\myto{\sigma_2}S_0)}{L} \myto{a.s.} \frac{P_{00}\prob{\sigma_2}}{\mu_{00}[\prob{\sigma_2}+\prob{\sigma_4}] }.
	$$
	
	The second and third term in $\alpha$ can be decomposed similarly:
	$$
	\begin{array}{rcl}
	\frac{\theta\#_L(S_0\myto{\sigma_4}S_0)}{L}&=&\theta\frac{\#_L S_0}{L} \cdot \frac{\#_L e_{00} }{\#_L S_0} \cdot \frac{\#_L(S_0\myto{\sigma_4} S_0)  }{\#_L e_{00}}\\
	&\myto{a.s.}& \frac{\theta P_{00}\prob{\sigma_4}}{\mu_{00}[\prob{\sigma_2}+\prob{\sigma_4}] },
	\end{array}
	$$
	and
	$$
	\begin{array}{rcl}
	\frac{\theta\#_L(S_1\myto{\sigma_4}S_0)}{L}&=&\theta\frac{\#_L S_1}{L} \cdot \frac{\#_L e_{10} }{\#_L S_1} \cdot \frac{\#_L(S_1\myto{\sigma_4} S_0)  }{\#_L e_{10}}\\
	&\myto{a.s.}& \frac{\theta P_{10}\prob{\sigma_4}}{\mu_{11}[\prob{\sigma_2}+\prob{\sigma_4}] }.
	\end{array}
	$$
	Hence, the existence of $\alpha$ has been proved, and the definition is well-defined. 
	We can compute $\alpha$ from this expression in terms of $p_h,\Delta,\theta$. 
	The theorem in \cite{sor} shows
	$$
	\frac{1}{\mu_{00}} = \frac{\pi_0}{\sum_{i=0}^2\pi_i\mu_i},
	\frac{1}{\mu_{11}} = \frac{\pi_1}{\sum_{i=0}^2\pi_i\mu_i}.
	$$
	Combining the results above, we get
	$$
  \begin{array}{rcl}
  \alpha &=& \frac{\pi_0 P_{00}\prob{\sigma_2}}{[\prob{\sigma_2}+\prob{\sigma_4}]\sum_{i=0}^2 \pi_i\mu_i}\left( 1- \frac{\pi_2P_{20}}{ \sum_{i=0}^2 \pi_i P_{i0} }  \right) \\
  & &+\theta \frac{\pi_0P_{00} \prob{\sigma_4} + \pi_1 P_{10}\prob{\sigma_4}   }{ [\prob{\sigma_2}+\prob{\sigma_4}]\sum_{i=0}^2 \pi_i\mu_i }
  \end{array}
  $$
	where $\mu_i$ denotes the expected time that the semi-Markov process stays in state $S_i$ before making a transition 
	and $\mu_i = \sum_{j=0}^2 P_{ij}l_{ij}$. 
\end{myproof}

\subsubsection{Proof of \cref{lem:concentration}}\hfill
\label{appendix: proof of concentration}

\begin{myproof}
Consider $k$ transitions in the Markov chain and $L$ denotes the total time slots of $k$ transitions. 
From the proof of \cref{well-defined alpha}, we can know that it suffices to prove the concentration bound for $\frac{\#_L(S_0\myto{\sigma_2}S_0)}{L} = \frac{k}{L} \cdot \frac{\#_L S_0 }{k} \cdot \frac{\#_L(S_0\myto{\sigma_2} S_0)  }{\#_L S_0}$.

By applying the Chernoff bound for $\frac{\#_L e_{00} }{\#_L S_0}$ and $\frac{\#_L(S_0\myto{\sigma_2} S_0)  }{\#_L e_{00}}$, we have
$$
\begin{array}{rcl}
& & \Prob{\frac{ \#_L(S_0\myto{\sigma_2} S_0)  }{\#_L S_0}  \leq \frac{ (1-\delta_1)P_{00} \prob{\sigma_2}}{\prob{\sigma_2}+\prob{\sigma_4}} }   \\
&\leq& exp({-c_1\delta_1^2\prob{\sigma_2} P_{00} /(\prob{\sigma_2}+\prob{\sigma_4})}),
\end{array}
$$
where $c_1$ is a constant independent of the random variables.

Let $\mathcal{K}$ be the $\epsilon$-mixing time for the embedded chain, with $\epsilon<\frac{1}{8}$. Applying the \cref{thm:Markov concentration}, we can obtain a concentration bound for visiting the state $S_0$:
$$
\prob{ \#_L S_0 \leq (1-\delta_2) k \pi_0   }\leq
c_2\mynorm{\phi}_{\pi} e^{-\delta_2^2 k\pi_0 / (72\mathcal{K})}.
$$

The time spent on each edge is a sub-exponential random variable. Let $Z_i$ be the i-th transition time so $L=\sum_{i=1}^kZ_i$ and
$$
\prob{L\geq (1+\delta_3) \sum_{i=1}^k\mathbb{E}_{\pi}{Z_i}     }  \leq c_3 e^{-c_4\delta_3 k},
$$
where $c_3,c_4$ are independent with $k$, and $ \mathbb{E}_{\pi}{Z_i} = \sum_{j=0}^2 \pi_j\mu_j$. 

Hence, conditioning on the event 
$$
\begin{array}{lcl}
\{ \#_L(S_0\myto{\sigma_2} S_0) \leq (1-\delta_1) \#_L S_0  \frac{ P_{00}\prob{\sigma_2}}{ {\prob{\sigma_2}+\prob{\sigma_4}}}  \}  
 \cup\{ \#_L S_0 \leq (1-\delta_2) k \pi_0   \} \cup \{  L\geq (1+\delta_3) \sum_{i=1}^k\mathbb{E}_{\pi}{Z_i}   \}  ,  
\end{array}
$$
choose $\delta$ such that $1-\delta = \frac{(1-\delta_1)(1-\delta_2)}{1+\delta_3}$ so 
$$
\alpha_L < (1-\delta)\alpha.
$$
Take the union bound:
$$
\prob{\alpha_L  < (1-\delta)\alpha} = \exp (-\Omega (\delta k)).
$$
The results shows that the probability of underestimating the consensus difficulty rate is negligible in $k$, the number of patterns in the given rounds.

In order to express this bound in terms of $L$, consider the concentration bound for $L$:
$$
k\Delta<L<(1+\delta_3)k\sum_{j=0}^2 \pi_j\mu_j
$$
with overwhelming probability in $k$. Hence, $L=\Theta(k)$ with high probability and
$$
\prob{\alpha_L  < (1-\delta)\alpha} = \exp (-\Omega (\delta L)).
$$
\end{myproof}
 
\subsubsection{Proof of \cref{lem: setting theta}}\hfill
\label{appendix: proof of low theta}

\begin{myproof}
  The probability for `h' to appear in each time slot is $q_h$
  Conditioning on pattern $\sigma_4$, the number of symbol `h' $Y$ follows a binomial distribution with parameters $\Delta$ and $q_h/(1-q_H)$. 
  We can obtain the inequality by applying the Chernoff's inequality to $Y$.
\end{myproof}

\subsubsection{Proof of \cref{prop:R>1}}\hfill
\label{appendix: proof of R}

\begin{myproof}
  The ratio 
  $$
  \begin{array}{rcl}
  \mathcal{R}  &=& \frac{(1-q_H)^\Delta }{p_h(q_h+\theta q_H)(1-p_h)^{2\Delta}}  \big[ q_h^2(1-p_h)^{2\Delta} 
+ q_hq_H(1-p_h)^\Delta(1-q_H)^\Delta+\theta p_hq_H(1-q_H)^\Delta \big].
  \end{array}
  $$
  which is a fractional linear function. 
  If we want to show the monotonicity of fractional linear function $f(x) = \frac{ax+b}{cx+d}$, 
  we only need to compare the values of $\frac{a}{c}$ and $\frac{b}{d}$,
  and $\frac{a}{c} <  \frac{b}{d}$ indicates the increasing property of $\mathcal{R}$ in $\theta$.

  It suffices to show $\frac{q_h^2(1-p_h)^{2\Delta} +q_hq_H (1-p_h)^\Delta (1-q_H)^\Delta  }{q_h} <  \frac{p_hq_H(1-q_H)^{\Delta}}{q_H}$:
  $$
  \begin{array}{rcl}
  & &\frac{q_h^2(1-p_h)^{2\Delta} +q_hq_H (1-p_h)^\Delta (1-q_H)^\Delta  }{q_h} \\
  &<& {q_h(1-p_h)^{2\Delta} +q_H (1-p_h)^\Delta}\\
  &<&   q_h(1-p_h)^{\Delta} +q_H (1-p_h)^\Delta \\
  &=& p_h(1-p_h)^{\Delta} \\
  &<& p_h(1-q_H)^{\Delta} \\
  &=& \frac{p_hq_H(1-q_H)^{\Delta}}{q_H}.
  \end{array}
  $$
  Therefore, $\mathcal{R}$ is increasing in $\theta$. Let $\theta=1$, and thus
  $$
  \begin{array}{lcl}
  \mathcal{R} > \frac{(1-q_H)^\Delta[q_h^2(1-p_h)^{2\Delta}+q_hq_H(1-q_H)^\Delta(1-p_h)^{\Delta}+p_hq_H(1-q_H)^{\Delta} ]}{p_h^2(1-p_h)^{2\Delta}}\\
  \ \ \ > \frac{(1-q_H)^\Delta[p_hq_h(1-p_h)^{2\Delta}+p_hq_H(1-q_H)^{\Delta} ]}{p_h^2(1-p_h)^{2\Delta}}\\
  \end{array}
  $$
  Apply the weighted power mean inequality to the RHS:
  $$
  \begin{array}{rcl}
   & & \frac{(1-q_H)^\Delta[p_hq_h(1-p_h)^{2\Delta}+p_hq_H(1-q_H)^{\Delta} ]}{p_h^2(1-p_h)^{2\Delta}} \\
   &=& \frac{q_h}{p_h}(1-q_H)^\Delta +\frac{q_H}{p_h}(\frac{1-q_H}{1-p_h})^{2\Delta} \\
   &\geq& \left[ \frac{q_h}{p_h}(1-q_H)+ \frac{q_H}{p_h} (\frac{1-q_H}{1-p_h})^2 \right]^\Delta\\
   &\geq& \left[ \frac{q_h}{p_h}(1-q_H)+ \frac{q_H}{p_h} (\frac{1-q_H}{1-p_h}) \right]^\Delta.
   \end{array}
  $$
  It remains to prove $\frac{q_h}{p_h}(1-q_H)+ \frac{q_H}{p_h}\left(\frac{1-q_H}{1-p_h}\right)\geq 1$ if we want to show $\mathcal{R}>1$. As
  $$
  \begin{array}{rcl}
    \frac{q_h}{p_h}(1-q_H)+ \frac{q_H}{p_h}\cdot \frac{1-q_H}{1-p_h} 
   &=&\frac{(1-q_H)(1-q_h)}{(1-p_h)}   \\
   &=& \frac{1-p_h +q_hq_H}{1-p_h} >1, \\
  \end{array}
  $$
  we conclude the proof.
\end{myproof}

\subsubsection{The Expression of $\mathcal{A}$ and Approximation Method}

The exact expression for the tolerance ratio is
$$
\begin{array}{rcl}
  \label{eq:expression of A}
\mathcal{A} &=& \frac{  (1-qp_h)^\Delta  }{(\theta -1)q+1}[ (1-p_h)^{2\Delta}(q-1)^2 \\
& & + (1-p_h)^\Delta q(1-q)(1-qp_h)^\Delta +\theta q(1-qp_h)^\Delta     ]. 
\end{array}
$$
Though $\mathcal{A}$ has an analytical expression, 
it is hard to obtain a closed-form expression of $q^*$. 
We can apply numerical methods such as Newton's iteration to find $q^*$.
In order to analyze the relationship between two parameters, 
we approximate the object function $\mathcal{A}$ by ignoring some terms with small values in $\mathcal{A}$.
Since $(1-p_h)^{2\Delta}(q-1)^2<(1-qp_h)^\Delta $ and $(1-p_h)^\Delta q(1-q)(1-qp_h)^\Delta < q(1-qp_h)^\Delta $, 
we can ignore these terms for large $\theta$ (e.g., $\theta>100$). 
Now, we can define the lower bound $\tilde{\mathcal{A}}$ for $\mathcal{A}$:
\begin{equation}
  \label{eq:A tilde}
\tilde{\mathcal{A}} :=  \frac{ \theta q (1-qp_h)^{2\Delta}  }{(\theta -1)q+1},
\end{equation}
which has a unique solution to the new optimization problem:
$$
q^* = \mathop{argmax}_{q\in (0,1)} \tilde{\mathcal{A}}.
$$

\begin{figure*}
  \centering
\begin{tabular}{rrr}

    \begin{tikzpicture}[scale=0.55]
    \begin{axis}[
        xlabel = $q$ (\qname),
        ylabel = $\mathcal{A}$ (Adversary tolerance ratio) 
    ]
    \addplot [mark=none, color=RoyalBlue] table [x index=1,y index=2, col sep = comma] {tex_figure/adv_ratio_q.csv};
    \end{axis}
    \end{tikzpicture}
&

  \begin{tikzpicture}[scale=0.55]
  \begin{axis}[
      xlabel = $\theta$ (weight parameter),
      xmax=5100,
      ylabel = $q^*$ (Optimal \qname),
      yticklabel style={  /pgf/number format/precision=3,    
                  /pgf/number format/fixed,    
                  /pgf/number format/fixed zerofill },
       xticklabel style={ 
                }
  ]

  \addplot [only marks, mark=halfcircle*, mark size=2.9pt,color = RoyalBlue ] table [x index=1,y index=2, col sep = comma] {tex_figure/qstar_theta.csv};
  \addplot  [color = RoyalBlue ]table [x index=1,y index=2, col sep = comma] {tex_figure/qstar_theta_approximate.csv};
  \end{axis}
  \end{tikzpicture}

&

  \begin{tikzpicture}[scale=0.55]
  \begin{axis}[
      xlabel = $\theta$ (weight parameter),
      xmax=5100,
      ylabel = $\mathcal{A}_\theta^*$ (Maximal adversary tolerance ratio),
      yticklabel style={  /pgf/number format/precision=3,    
                  /pgf/number format/fixed,    
                  /pgf/number format/fixed zerofill },
       xticklabel style={ 
                }
  ]

  \addplot [only marks, mark=halfcircle*, mark size=2.9pt,color = RoyalBlue ]  table [x index=1,y index=2, col sep = comma] {tex_figure/theta_opt_adv.csv};
  \addplot [color = RoyalBlue ]table [x index=1,y index=2, col sep = comma] {tex_figure/theta_opt_adv_approx.csv};
  \end{axis}
  \end{tikzpicture}

\end{tabular}
\centering
\caption{$\Delta=10^{13}$, $p_h=7.5\times 10^{-14}$ in each figure. 
The first figure is a theoretical curve of adversary tolerance ratio $\mathcal{A}$ with the case $\theta=100$. 
The second figure shows the decreasing trend of optimal $q$ in $\theta$. 
The third figure is the maximal adversary tolerance ratio for each $\theta$.
The curves in the second and third graphs are results of approximation method 
while the sampled points are obtained by maximizing $\mathcal{A}$.
}
\label{fig:adv_ratio in theory}
\end{figure*}
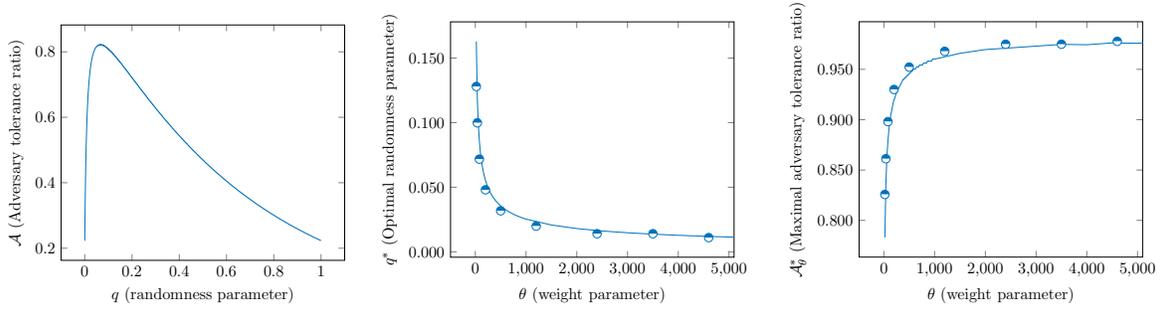


\subsubsection{Proof of \cref{prop:adversary ratio properties}}\hfill
\label{appendix: proof of A}

\begin{myproof}
    1. Regard $\mathcal{A}$ as the single variable function of $q$. As $\mathcal{A}$ is continuous in $[0,1]$, 
    it must have a maximizer in $[0,1]$. It suffices to show
    $\frac{\partial \mathcal{A}}{\partial q} \big{|}_{q=1}<0$ and $\frac{\partial \mathcal{A}}{\partial q} \big{|}_{q=0}>0$ to rule out the case that the maximizer is the endpoint of $[0,1]$.
    
    Then 
    $$
    \frac{\partial \mathcal{A}}{\partial q} \bigg{|}_{q=0} = \theta[1-(1-p_h)^{2\Delta}] +(1-p_h)^\Delta[ 1- (1+\Delta p_h) (1-p_h)^{\Delta}]
    $$
    is positive and
    $$
    \frac{\partial \mathcal{A}}{\partial q} \bigg{|}_{q=1} = \frac{1}{\theta}(1-p_h)^{2\Delta-1}[ -2\Delta p_h\theta + (1-p_h)(1-(1-p_h)^\Delta)   ]
    $$
    is negative.
    
    Clearly, $\theta[1-(1-p_h)^{2\Delta}]>0$ and $1- (1+\Delta p_h) (1-p_h)^{\Delta}\geq 0$ 
    as a result of the function $f(x)= (1+nx)(1-x)^n\leq 1$ for $x\in [0,1]$ and $n\geq 1$. 
    Hence, $\frac{\partial \mathcal{A}}{\partial q} \big{|}_{q=0}>0$. 

    For the second inequality, the Bernoulli inequality shows that $(1-p_h)^\Delta>1-\Delta p_h$. 
    Combining the fact that $\theta>1$, we obtain
    $$
    \frac{\partial \mathcal{A}}{\partial q} \bigg{|}_{q=1} < -\frac{1}{\theta}(1-p_h)^{2\Delta-1} (1+p_h)p_h\Delta  < 0
    $$
    
    Since the derivative $\frac{\partial \mathcal{A}}{\partial q}$ is continuous in $(0,1)$, 
    the function will have a nontrivial maximizer in $(0,1)$.

    2. Let $\frac{\partial \tilde{\mathcal{A}}}{ \partial q} =0$ and solving the equation,
    we obtain 
    $$
    q^*(\theta) = \frac{-{p_h}({2\Delta +1}) + \sqrt{p_h^2 (2\Delta+1)^2  + 8\Delta p_h (\theta-1) }}{4\Delta{p_h}(\theta - 1)}.
    $$
    The monotonicity of $q^*(\theta)$ follows directly from the above expression.
    
    3. It is observable that $\mathcal{A}$ is increasing in $\theta$.
    Therefore, it is straightforward that $\mathcal{A}^*_{\theta}$ is increasing in $\theta$.        
\end{myproof}

\subsection{Chain Properties}
\subsubsection{Proof of \cref{lem:chain growth}}\hfill
\label{appendix:chain growth}

\begin{myproof}
  At the time slot $t$, each honest player will try to extend a path whose weight is at least $\diffFC{t}$, 
  because at time $t-\Delta$, the heaviest path has weight $\diffFC{t}$ and after $\Delta$ rounds, 
  every honest party will hold a chain with weight at least $\diffFC{t}$. 

  Let $\tau$ denote the time interval such that 
  $$
  \tau = \min\{t':\diffFC{t+t'}\geq \diffFC{t}+(1-q)+q\theta \},
  $$
  then $\e{\tau} \leq \frac{1}{p_h}+\Delta$. The expected weight of a block is $(1-q)+q\theta$, and 
  the expected time to produce a block for honest parties is $\frac{1}{p_h}$. Therefore, after $\Delta$ time slots, 
  each honest party will hold a chain with weight at least $\diffFC{t+\tau}$ at time $t+\tau$. 
  Then we conclude that in the sense of expectation, 
  one will extend a block to the heaviest chain in $F_t$ with time interval at most $\frac{1}{p_h}+\Delta$. 
  Then for $S$ rounds, the heaviest chain will increase weight at least 
  $$
  S p_h\frac{1-q+q\theta}{1+p_h\Delta},
  $$
  and this indicates $\diffFC{t+S}\geq S p_h\frac{1-q+q\theta}{1+p_h\Delta}$ in the sense of expectation. 
  We finish the proof by applying the Hoeffding's inequality:
  $$
  \diffFC{t+S}-\diffFC{t}\geq (1-\delta)S p_h\frac{1-q+q\theta}{1+p_h\Delta}
  $$
  with probability $1-e^{\Omega(-S{\delta}^2)}$.
\end{myproof}
  
\begin{myrem}
    \label{rem: loose chain growth rate}
    From the proof of the \cref{lem:chain growth}, we can see the bound for $\e{\tau}$ is loose for some cases. 
    When $p_h\approx 1$, $\e{\tau}$ is much less than $\Delta$. 
    Therefore, the chain growth rate derived in the \cref{lem:chain growth} is not tight.
    Theoretically, the chain growth rate $u$ should be larger than the consensus rate $\alpha$, 
    because there are a proportion of weights from the blocks which do not lead to consensus in the heaviest chain.
    However, $u$ is less than $\alpha$ for specific parameters, 
    e.g., $p_h=2\times 10^{-13},\Delta = 10^{13},q=0.02,\theta=500$.
\end{myrem}

\begin{mylem} \label{lem:adv upper bound}
  Let $D_t$ denote the cumulative weights of blocks mined by adversary parties during $t$ time slots, then for any $\delta>0$,
  $$
  \prob{D_t >(1+\delta)p_at(1-q+q\theta) }\leq \exp(-\Omega(\delta^2t)).
  $$
\end{mylem}
\begin{myproof}
  WLOG, we can assume the beginning time slot is $1$. 
  Let random variable $X_i$ denote the weights contributed by the adversary at slot $i\in[t]$. 
  The probability mass function of $X_i$ is 
  $$
  \prob{X_i=1}=q_a , \prob{X_i=\theta}=q_A, \prob{X_i=0}=1-p_a.
  $$
  Since $X_i$ is a bounded random variable, then
  $$
  \begin{array}{rcl}
  & & \prob{D_t\geq(1+\delta)t p_a(1-q+q\theta)}\\&=&\prob{\sum_{i=1}^t X_i \geq(1+\delta)t \e{X_i}}\\ &\leq& \exp({-\Omega(\delta^2t)}),
  \end{array}
  $$
  where the last inequality is directly from the Hoeffding's inequality.
\end{myproof}

\subsubsection{Proof of \cref{lem:chain quality}}
\label{appendix:chain quality}

  Suppose $\mathcal{S}\subset \chain_{t-\Delta}$ starts with an honest block $B$ whose label is $t_1$ 
  and the cumulative weight of $\mathcal{S}$ is $D$, then it suffices to show that 
  the weight contributed by adversary blocks is at most $(1+\delta)(\frac{p_a}{p_h}+p_a \Delta)D$ in $\mathcal{S}$ 
  with high probability for every $\delta>0$. 
  
  Since $B$ is honest, then we have $\diff{\chain_{t_1}} \leq \diff{ \overline{ F_{t_1}} }$ 
  according to the definition of the maximal honest subfork. 
  Thus, by Lemma~\ref{appendix:chain growth}, for every $\delta_1>0$, 
  the chain growth rate of the this segment is at least $(1-\delta_1)p_h \frac{1-q+q\theta}{1+p_h\Delta}$ 
  except with probability $\exp({-\Omega(\delta_1^2D)})$. Then
  $$
  t-\Delta-t_1\leq \frac{D(1+p_h\Delta)}{(1-\delta_1)(1-q+q\theta)p_h}.
  $$
  
  By \cref{lem:adv upper bound}, for every $\delta_2>0$, 
  the weights contributed by adversary blocks in the time interval are upper bounded by 
  $$
  (1+\delta_2)p_a (t-\Delta-t_1)(1-q+q\theta) \leq \frac{(1+\delta_2)(\frac{p_a}{p_h}+p_a\Delta) D}{(1-\delta_1)}
  $$
  except with probability $e^{-\Omega(\delta_1^2D)}+e^{-\Omega(\frac{\delta_2^2}{1-\delta_1}D)}$.
  
  For a given $\delta >0$, we can select a $\delta_1$ and let $1+\delta = \frac{1+\delta_2}{1-\delta_1}$, 
  so the fraction of the cumulative weights contributed by adversary parties are at most 
  $(1+\delta)(\frac{p_a}{p_h}+p_a \Delta)$ except with probability $\exp({-\Omega(\delta^2D)})$.

 
  

\subsection{Related Material}

\begin{mythm}[\cite{markov_concentration}]
\label{thm:Markov concentration}
Let $M$ be an ergodic Markov chain with state space $[n]$ and stationary distribution $\pi$. Let $\mathcal{T}$ be its $\epsilon$-mixing time for $\epsilon<1/8$. Let $(V_1,V_2,\cdots,V_t)$ denote a $t$-step random walk on $M$ staring from an initial distribution $\phi$ on $[n]$. For every $i\in [t]$, let $f_i:[n]\to [0,1]$ be a weighted function at step $i$ such that the expected weight $\mathbb{E}_{\pi}[f_i(v)] = \mu$ for all $i$. Define the total weight of the walk $(V_1,V_2,\cdots,V_t)$ by $X=\sum_{i=1}^t f_i(V_i)$. There exists some constant $c$ (which is independent of $\mu$,$\delta$, and $\epsilon$) such that
\begin{enumerate}
    \item $\prob{X\geq (1+\delta)\mu t }\leq c\mynorm{\phi}_\pi exp (-\delta^2\mu t / (72\mathcal{T}))$ for $0 \leq \delta \leq 1$,
    \item $\prob{X\geq (1+\delta)\mu t }\leq c\mynorm{\phi}_\pi exp (-\delta\mu t / (72\mathcal{T}))$ for $ \delta > 1$,
    \item $\prob{X\leq (1-\delta)\mu t }\leq c\mynorm{\phi}_\pi exp (-\delta^2\mu t / (72\mathcal{T}))$ for $0\leq \delta \leq 1$,
\end{enumerate}
 where $\mynorm{\phi}_\pi $ is the $\pi$-norm of $\phi$ given by $\sum_{i\in [n]} \phi_i^2 /\pi_i $.
\end{mythm}

\begin{mylem}
  \label{lem:ineq1}
$f(x) = (1+nx)(1-x)^n\leq 1$ for $x\in[0,1]$ and $n\geq 1$.
\end{mylem}

\begin{myproof}
    $$
    f'(x) = - n(n+1)x(1-x)^{n-1}\leq 0 ,
    $$
so $f(x)\leq f(0)=1$ for $n>1$. As for $n=1$, $f(x)=1-x^2<1$.
 
\end{myproof}

\begin{mylem}[Weighted Power Mean Inequality]
  \label{lem:powermean}
Assume $a_i$, $b_i$ and $p_i$, $ i=1,2,\cdots,k$ are positive numbers. For $m>n$, then
$$
\left(\frac{\sum_{i=1}^k p_ia_i^m }{\sum_{i=1}^k p_i}  \right)^{\frac{1}{m}}\geq 
\left(\frac{\sum_{i=1}^k p_ia_i^n }{\sum_{i=1}^k p_i}  \right)^{\frac{1}{n}}.
$$
The equality holds iff $a_i=b_i$ for $i=1,2,\cdots,k$.
\end{mylem}

\begin{mythm} [McDiarmid’s Inequality]
  Let $X_1, \cdots , X_n$ be independent random variables, where $X_i$ has range $\mathcal{X}_i$. 
  Let $f : \mathcal{X}_1 \times \cdots \times  \mathcal{X}_n \to \mathbb{R}$ be any function
  with the $(c_1,\cdots , c_n)$-bounded differences property: for every $i = 1, \cdots , n$ and
  every $(x_1, \cdots , x_n),(x'_1, \cdots , x'_n) \in \mathcal{X}_1 \times \cdots \times  \mathcal{X}_n$ that differ only in the $i$-th
  corodinate ($x_j = x'_j$ for all $j\neq i$ ), we have
  $$
  |f(x_1,\cdots,x_n)-f(x'_1,\cdots,x'_n)| \leq c_i.
  $$
  For any $t>0$,
  $$
  \begin{array}{rcl}
  & &\prob{ f(X_1,\cdots,X_n)  - \e{ f(X_1,\cdots,X_n)}\geq t  } \\
  &\leq& \exp \left( -\frac{2t^2}{\sum_{i=1}^n c_i^2}   \right).
  \end{array}
  $$
\end{mythm}